# Dynamic Response and Stability Margin Improvement of Wireless Power Receiver Systems via Right-Half-Plane Zero Elimination

Kerui Li, *Student Member, IEEE*, Siew-Chong Tan, *Senior Member, IEEE,* and Ron Shu Yuen Hui, *Fellow, IEEE*

*Abstract-* **The series-series compensation topology is widely adopted in many wireless power transfer applications. For such systems, their wireless power receiver part typically involves a DC-DC converter with front-stage full-bridge diode rectifier, to process the high-frequency transmitted AC power into a DC output voltage for the load. It is recently reported that the current source nature of the series-series compensation will introduce right-half-plane (RHP) zeros into the small-signal transfer functions of the DC-DC converter of the wireless power receiver, which will severely affect the stability and dynamic response of the system. To resolve this issue, in this paper, it is proposed to adopt a different rectifier configuration for the system such that the input current to the DC-DC converter becomes controllable to eliminate the presence of RHP zeros of the small-signal transfer functions of the system. This rectifier can be applied to different wireless power receivers using the buck, buck-boost, or boost converters. As compared with the original wireless power receivers, the modified ones feature minimum-phase characteristics and hence ease the design of compensator. Theoretical and experimental results are provided. The comparative experimental results verify the elimination of the RHP zero, improved dynamic responses of reference tracking and against load disturbances, and a larger stability margin.**

## I. INTRODUCTION

The wireless power receivers which comprise a front-end diode rectifier and a post DC-DC converter [1], [2], are popular solutions for wireless charging applications. Since the rectified DC voltage provided by the diode rectifier does not necessarily match the voltage requirement of the load, the DC-DC converters are hence utilized to match the voltage difference and regulate the output voltage [1], [3].

The charging profile of the battery imposes strict requirements on the charging current and voltage [4], which, in turn, raises new challenges on the output voltage/current regulation of the wireless power receivers [5], [6]. Despite many research efforts on improving the regulation capability of the DC-DC converters, these works are not necessarily applicable to those DC-DC converters used in wireless power receivers. Conventional DC-DC converters are designed under the condition of voltage source input. However, the input of the DC-DC converters employed in wireless power receivers is of current source nature when adopting the series-series compensation [7], [8]. This current source is non-controllable under the existing full-bridge diode rectifier structure, and it alters both the large-signal and small-signal characteristics of the DC-DC converter system. It leads to highly load-dependent steady-state DC-link and output voltages [9]. Additionally, the finite DC-link capacitance introduces a non-negligible phase delay to the small-signal response of the DC-link voltage and the phase delay further adversely affects the dynamic characteristics of the DC-DC converter [10]. Therefore, the small-signal response of the output voltage is coupled with the DC-link capacitance and shows extra phase delay [11]. A right-half-plane (RHP) zero is thereby introduced to the duty ratio-to-output voltage transfer function [12]. For example, the buck converter, which was considered a minimum-phase (no RHP zero) system, is changed into a non-minimum-phase system when it is utilized in the wireless power receiver under the series-series compensated configuration [12]. The design of the compensator becomes more challenging in the presence of RHP zeros [13]-[15]. The stability margin and dynamic response of the receiver are adversely affected [16], [17]. To improve the performance of the wireless power receivers, various control methods, e.g., the dual-side controller [1], the dual-loop controller [12], and the nonlinear controllers [18], [19] have been proposed. However, for the control method reported in [1], the performance of the receivers is still relatively poor as it is highly dependent on the length of wireless communication latency, of which neither a fast dynamic response nor large stability margin can be attained. For the control method reported in [12], the stability margin remains limited. For the nonlinear control methods reported in [18] and [19], they suffer from complicated design procedures and high computation burden. These methods do not address the non-controllable input current source nature of the DC-DC converter. Therefore, RHP zeros are present in these systems and remain the limiting factor confining their stability margin and dynamic performance.

In this work, the RHP zeros of wireless power receivers using different DC-DC converters are investigated. The full-bridge rectifier adopts a different configuration, which enables certain control of the AC input current to the DC-DC converter. In doing so, RHP zeros can be subsequently eliminated, thereby improving the closed-loop bandwidth and stability margin of the wireless power receiver system.

## II. RHP ZEROS OF THE WIRELESS POWER RECEIVERS

### *A. Wireless Power Receivers*

Fig. 1 shows the wireless power transfer systems with the various possible types of wireless receivers. On the transmitter side, a full-bridge inverter operating at switching frequency $f$ ($f$=1/$T$) and a primary-side transmitter coil $L_p$ with a series compensation capacitor $C_p$ (i.e., $1/\sqrt{C_pL_p} = 2\pi f$) are used. On the receiver side, the secondary-side receiver coil $L_s$ with a series compensation capacitor $C_s$ (i.e., $1/\sqrt{C_s L_s} = 2\pi f$), a passive diode rectifier ($D_1$ to $D_4$) for rectification, and a second-stage DC-DC converter for matching the voltage of the load are used. The wireless power receivers with typical DC-DC converters (i.e. the buck converter, buck-boost converter, and boost converter) are shown in Fig. 1 (a)-(c) respectively.

Due to the series-series compensation, the wireless power receiver coil $L_s$ introduces an independent sinusoidal current $i_{Ls}(t)$ to the diode rectifier [7], [8], i.e.,

$$i_{Ls}(t) = I_{Ls}\sin(2\pi f t), \tag{1}$$

where $I_{Ls}$ is the amplitude of the sinusoidal current $i_{Ls}(t)$. The independent current source nature and finite DC-link capacitor introduce RHP zeros to the second-stage DC-DC converters.

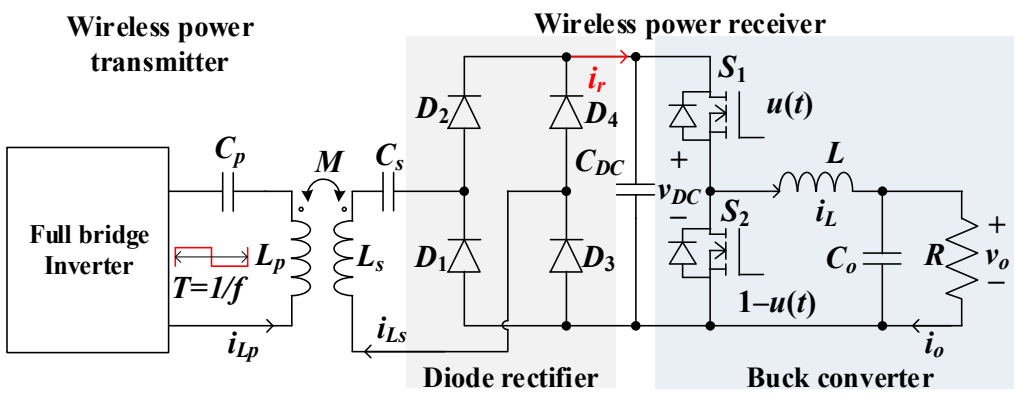

(a) With full-bridge diode rectifiers and buck converter

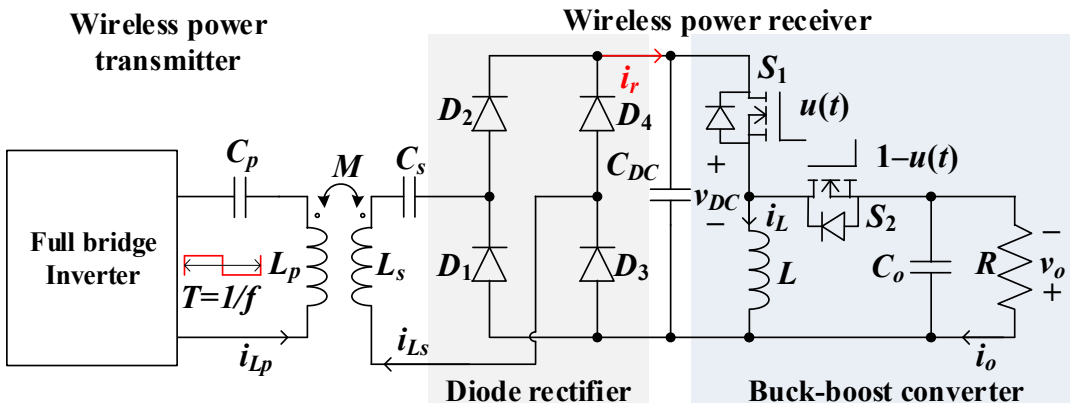

(b) With full-bridge diode rectifiers and buck-boost converter

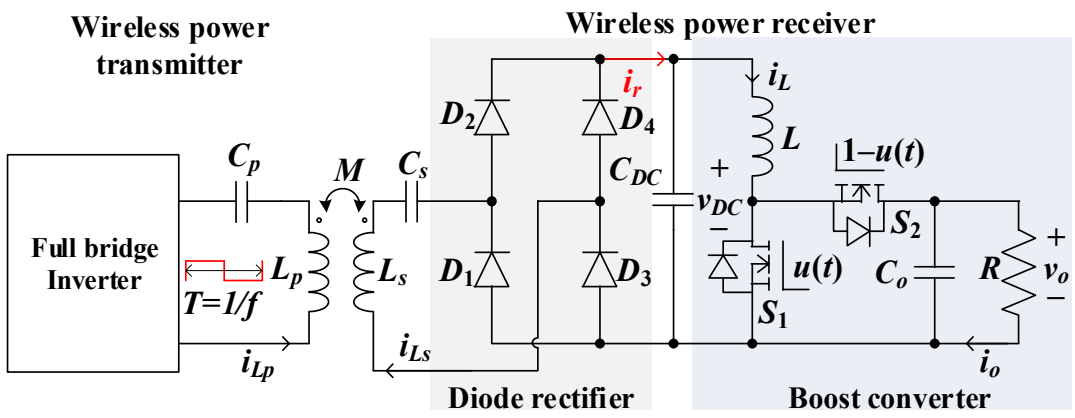

(c) With full-bridge diode rectifiers and boost converter

Fig. 1. Wireless power transfer systems with various receivers.

## *B. Transfer Functions and The RHP Zeros*

To facilitate the analysis, the DC-DC converters share identical notations i.e., complementary MOSFET switches $S_1$ and $S_2$, inductor $L$, output capacitor $C_o$ and load $R$. The switching functions of the switch $S_1$ and the complementary switch $S_2$ are $u(t)$ and $1-u(t)$, respectively. The switching function $u(t)$ is written as

$$u(t) = \begin{cases} 1 & if\ nT < t \le (n + D_{DC-DC})T \\ 0 & if\ (n + D_{DC-DC})T < t \le (n+1)T \end{cases} \tag{2}$$

where logic 1 and 0 represent the ON and OFF state of the switch. Particularly, $D_{DC-DC}$ represents the duty ratio of $u(t)$ over one switching period $T$, and $n$ is an arbitrary non-negative integer. In addition, to avoid the beat frequency oscillation and ease of the analysis, the switching frequency of $u(t)$ is synchronized at $f$, in a cycle-by-cycle manner [20].

*(i) Wireless power receiver with buck converter*

The state-space equation of the wireless power receiver with buck converter is written as

$$\begin{bmatrix} \frac{\partial\langle v_{DC}\rangle_T}{\partial t} \\ \frac{\partial\langle i_L\rangle_T}{\partial t} \\ \frac{\partial\langle v_o\rangle_T}{\partial t} \end{bmatrix} = \begin{bmatrix} 0 & -\frac{\mathbf{D_{DC-DC}}}{C_{DC}} & 0 \\ \frac{\mathbf{D_{DC-DC}}}{L} & 0 & -\frac{1}{L} \\ 0 & \frac{1}{C_o} & -\frac{1}{RC_o} \end{bmatrix} \begin{bmatrix} \langle v_{DC}(t)\rangle_T \\ \langle i_L\rangle_T \\ \langle v_o\rangle_T \end{bmatrix} + \begin{bmatrix} \frac{2I_{Ls}}{\pi C_{DC}} \\ 0 \\ 0 \end{bmatrix} \tag{3}$$

where $<v_{DC}>_T$, $<i_L>_T$, and $<v_o>_T$ respectively represent the average of $v_{DC}(t)$, $i_L(t)$, and $v_o(t)$ over one switching cycle $T$. The control variable $D_{DC-DC}$ of the DC-DC converter is highlighted. Via small-signal linearization with respect to the control variable $D_{DC-DC}$ [16], the control-to-output voltage transfer function is

$$G_{buck}(s) = \frac{2RI_{Ls}(C_{DC}Rs - D_{DC-DC}{}^2)}{\pi D_{DC-DC}{}^2[C_oC_{DC}LRs^3 + C_{DC}Ls^2 + (C_oRD_{DC-DC}{}^2 + C_{DC}R)s + D_{DC-DC}{}^2]} \tag{4}$$

There is one RHP zero in the transfer function, which is

$$Z_{buck} = \frac{D_{DC-DC}{}^2}{C_{DC}R} \tag{5}$$

*(ii) Wireless power receiver with buck-boost converter*

Likewise, the state-space model of the wireless power receiver with buck-boost converter is derived as

$$\begin{bmatrix} \frac{\partial\langle v_{DC}\rangle_T}{\partial t} \\ \frac{\partial\langle i_L\rangle_T}{\partial t} \\ \frac{\partial\langle v_o\rangle_T}{\partial t} \end{bmatrix} = \begin{bmatrix} 0 & -\frac{\mathbf{D_{DC-DC}}}{C_{DC}} & 0 \\ \frac{\mathbf{D_{DC-DC}}}{L} & 0 & \frac{\mathbf{D_{DC-DC}} - 1}{L} \\ 0 & \frac{1-\mathbf{D_{DC-DC}}}{C_o} & -\frac{1}{RC_o} \end{bmatrix} \begin{bmatrix} \langle v_{DC}(t)\rangle_T \\ \langle i_L\rangle_T \\ \langle v_o\rangle_T \end{bmatrix} + \begin{bmatrix} \frac{2I_{Ls}}{\pi C_{DC}} \\ 0 \\ 0 \end{bmatrix} \tag{6}$$

Its control-to-output transfer function can be derived as

$$G_{buck-boost}(s) = \frac{-2RI_{Ls}(C_{DC}LD_{DC-DC}s^2 - C_{DC}R(1-D_{DC-DC})^2s + D_{DC-DC}{}^2)}{\pi D_{DC-DC}{}^2[C_oC_{DC}LRs^3 + C_{DC}Ls^2 + (C_oD_{DC-DC}{}^2 + C_{DC}(1-D_{DC-DC})^2)Rs + D_{DC-DC}{}^2]} \tag{7}$$

There are two RHP zeros in the transfer function, which are

$$Z_{buck-boost,1} = \frac{C_{DC}R(1-D_{DC-DC})^2 + \sqrt{[C_{DC}R(1-D_{DC-DC})^2]^2 - 4(D_{DC-DC})^3C_{DC}L}}{2C_{DC}LD_{DC-DC}}$$

$$Z_{buck-boost,1} = \frac{C_{DC}R(1-D_{DC-DC})^2 - \sqrt{[C_{DC}R(1-D_{DC-DC})^2]^2 - 4(D_{DC-DC})^3C_{DC}L}}{2C_{DC}LD_{DC-DC}} \tag{8}$$

*(iii) Wireless power receiver with boost converter*

The state-space model of the wireless power receiver with boost converter is

$$\begin{bmatrix} \frac{\partial\langle v_{DC}\rangle_T}{\partial t} \\ \frac{\partial\langle i_L\rangle_T}{\partial t} \\ \frac{\partial\langle v_o\rangle_T}{\partial t} \end{bmatrix} = \begin{bmatrix} 0 & -\frac{1}{C_{DC}} & 0 \\ \frac{1}{L} & 0 & -\frac{1-\mathbf{D_{DC-DC}}}{L} \\ 0 & \frac{1-\mathbf{D_{DC-DC}}}{C_o} & -\frac{1}{RC_o} \end{bmatrix} \begin{bmatrix} \langle v_{DC}(t)\rangle_T \\ \langle i_L\rangle_T \\ \langle v_o\rangle_T \end{bmatrix} + \begin{bmatrix} \frac{2I_{Ls}}{\pi C_{DC}} \\ 0 \\ 0 \end{bmatrix} \tag{9}$$

Its control-to-output transfer function can be derived as

$$G_{boost}(s) = \frac{-2RI_{Ls}(C_{DC}Ls^2 - C_{DC}R(1-D_{DC-DC})^2 s + 1)}{\pi[C_o C_{DC} LRs^3 + C_{DC}Ls^2 + (C_o + C_{DC}(1-D_{DC-DC})^2)Rs + 1]} \quad (10)$$

Here, there are two RHP zeros, which are

$$Z_{boost,1} = \frac{C_{DC}R(1-D_{DC-DC})^2 + \sqrt{[C_{DC}R(1-D_{DC-DC})^2]^2 - 4C_{DC}L}}{2C_{DC}L}$$

$$Z_{boost,2} = \frac{C_{DC}R(1-D_{DC-DC})^2 - \sqrt{[C_{DC}R(1-D_{DC-DC})^2]^2 - 4C_{DC}L}}{2C_{DC}L} \quad (11)$$

The physical origin of the RHP zeros is attributed to the non-controllable nature of diode rectifier and the current source input due to series-series compensation. As the load power changes, the DC-DC converter must adjust its duty ratio to meet the requirement of the load. However, as the input current to the receiver is non-controllable by the converter itself (and can only be controlled at the transmitter side), the converter cannot immediately adjust the input power to meet the load power requirement. This causes a temporary power imbalance and phase delay in the output voltage response. The temporary power imbalance and phase delay in output voltage response are the origins of the RHP zeros [12]. They impose fundamental limitations to the bandwidth of the closed-loop system, and adversely affect the dynamic response and stability margin [16], [17]. To ensure sufficient phase and gain margin, the closed-loop bandwidth must be reduced, which thereby results in a slower dynamic response of the receiver. Moreover, due to the high-frequency amplification nature of RHP zeros, high-frequency disturbances may easily generate oscillation and may severely affect the system's stability.

## III. Circuit Modifications and RHP Zeros Elimination

### *A. Modified Wireless Power Receiver with Active Rectifier*

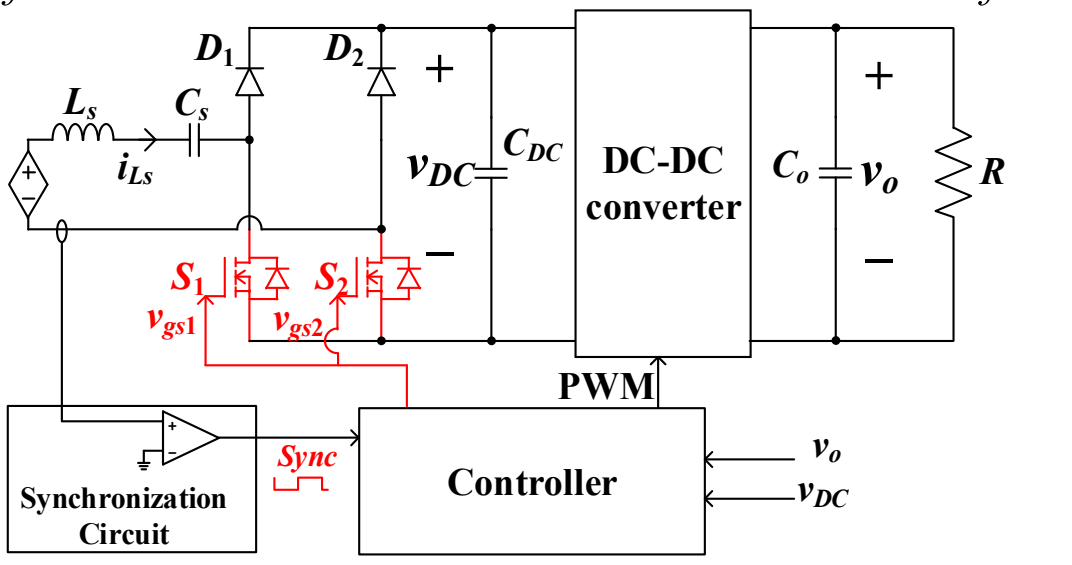

Fig. 2. The modified wireless power receiver.

The proposed modified rectifier enables output regulation to be more easily attainable via the control of the input current. It is through the proper control of the input current that the RHP zeros can be eliminated. Various types of active rectifiers [21]-[23] can be adopted to enable the function of AC input current control of the receiver. Considering the popularity of the full-bridge diode rectifier, it is adopted in this work the two ground-referenced active switches type of full-bridge diode rectifier, as shown in Fig. 2, for the investigation. An external synchronization circuit is used to convert the AC current $i_{Ls}$ into a square wave for the controller [20]. To reduce switching loss, the rising edge of the gate driving signals are aligned with the zero crossing points of $i_{Ls}$, and thereby the switches $S_1$ and $S_2$ can operate with zero voltage switching (ZVS) with reduced switching loss. In addition, the introduced switches cause less conduction loss as compared with diodes. The modification will not deteriorate the efficiency of the receiver system. Moreover, the cascaded DC-DC converter can operate with fixed duty ratio to match the voltage difference between the DC-link voltage and output voltage, which may further increase the efficiency of the DC-DC converter. Both the duty ratios of the gate driving signal $v_{gs1}$ and $v_{gs2}$ are $D$ ($0.5 \le D \le 1$), and they are 180-degree phase shift with one another. Fig. 3 and Fig. 4 respectively show the equivalent circuit models and key forms in different operation states.

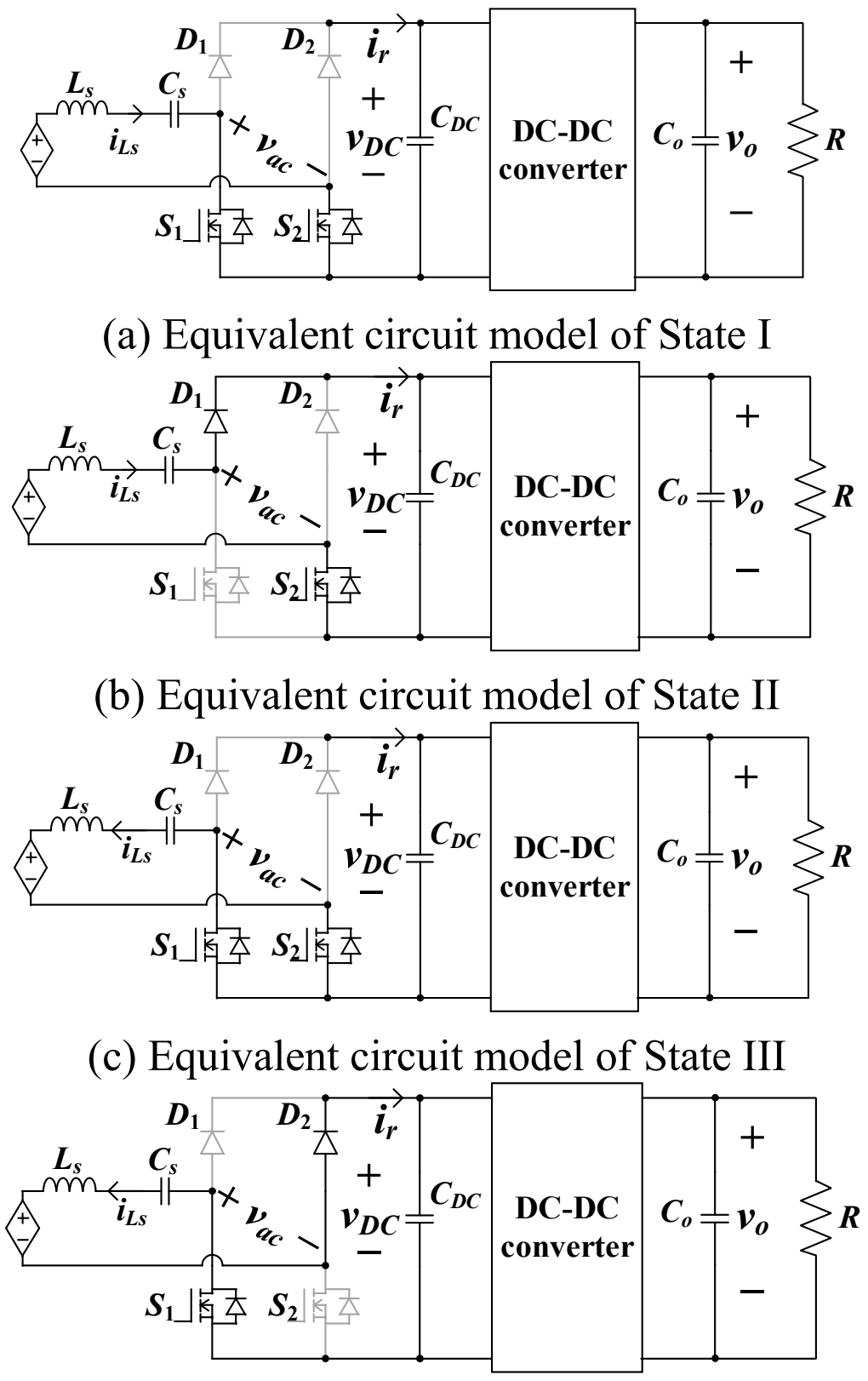

(a) Equivalent circuit model of State I

(b) Equivalent circuit model of State II

(c) Equivalent circuit model of State III

(d) Equivalent circuit model of State IV

Fig. 3. Equivalent circuit models of the modified wireless power receiver in different operation states.

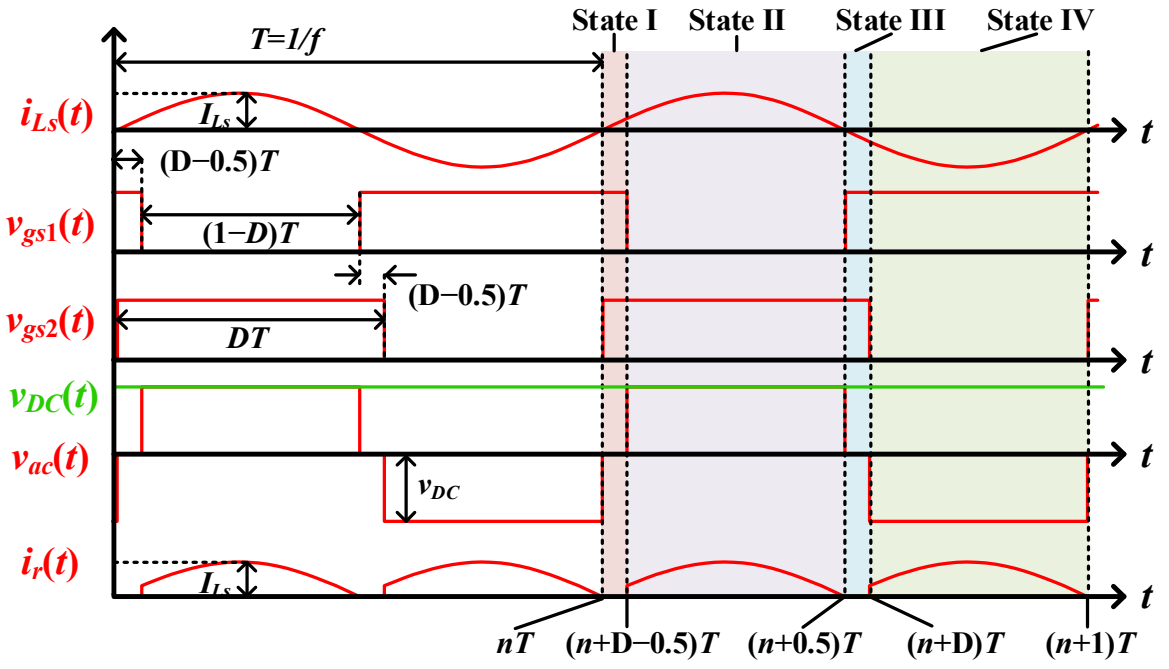

Fig. 4. Key waveforms of the wireless power receiver system.

*State* I $[nT \le t < (n+D-0.5)T]$:

Fig. 3(a) shows the equivalent circuit model of the power circuit in State I. At $t=nT$, $S_2$ turns ON with ZVS. During State I, both switches are ON. The AC current $i_{Ls}$ is freewheeled and no current flows to the DC-link capacitor.

*State* II $[(n+D-0.5)T \le t < (n+0.5)T]$:

Fig. 3(b) shows the equivalent circuit model of the power circuit in State II. At $t=(n+D-0.5)T$, $S_1$ turns OFF and $D_1$ turns ON. During State II, $i_{Ls}(t)$ flows to the DC-link capacitor.

*State* III $[(n+0.5)T \le t < (n+D)T]$:

Fig. 3(c) shows the equivalent circuit model of the power circuit in State III. At $t=(n+0.5)T$, $S_1$ turns ON with ZVS. During State III, both switches are ON. $i_{Ls}$ is freewheeled and no current flows to the DC-link capacitor.

*State* IV $[(n+D)T \le t < (n+1)T]$:

Fig. 3(d) shows the equivalent circuit model of the power circuit in State III. At $t=(n+D)T$, $S_2$ turns OFF and e $D_1$ turns ON. During State IV, $i_{Ls}(t)$ flows to the DC-link capacitor.

In general, the average current $\langle i_r \rangle_T$ flows to the DC-link capacitor can be obtained as

$$\langle i_r \rangle_T = \frac{1}{T}\int_{nT}^{(n+1)T} i_r(t)dt = \frac{I_{Ls}}{\pi}(1-\cos(2\pi D)) \qquad (12)$$

*B. Averaged Model and Small-Signal Model*

*(i) Modified wireless power receiver with buck converter*

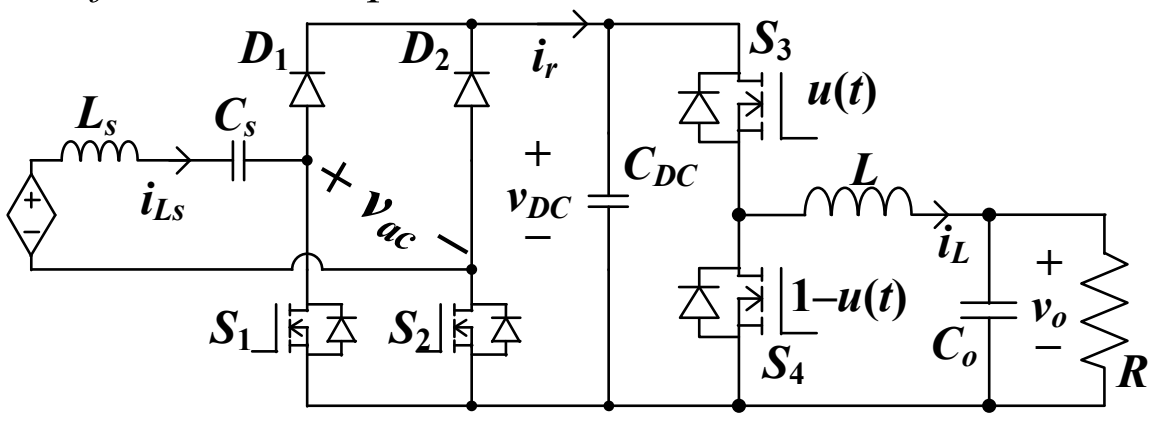

Fig. 5. Modified wireless power receiver with buck converter.

Fig. 5 shows the schematic diagram of the modified wireless power receiver with buck converter. The state-space averaged model is derived as

$$\begin{bmatrix} \frac{\partial \langle v_{DC} \rangle_T}{\partial t} \\ \frac{\partial \langle i_L \rangle_T}{\partial t} \\ \frac{\partial \langle v_o \rangle_T}{\partial t} \end{bmatrix} = \begin{bmatrix} 0 & -\frac{D_{DC-DC}}{C_{DC}} & 0 \\ \frac{D_{DC-DC}}{L} & 0 & -\frac{1}{L} \\ 0 & \frac{1}{C_o} & -\frac{1}{RC_o} \end{bmatrix} \begin{bmatrix} \langle v_{DC}(t) \rangle_T \\ \langle i_L \rangle_T \\ \langle v_o \rangle_T \end{bmatrix} + \begin{bmatrix} \frac{[1-\cos(2\pi \boldsymbol{D})]I_{Ls}}{\pi C_{DC}} \\ 0 \\ 0 \end{bmatrix} \qquad (13)$$

The control variable of the active rectifier $D$ is in bold. The control-to-output voltage transfer function of the modified wireless power receiver with buck converter is

$$G_{M_{buck}}(s) = \frac{2D_{DC-DC}I_{Ls}R\sin(2\pi D)}{\left[C_oC_{DC}LRs^3 + C_{DC}Ls^2 + (C_oRD_{DC-DC}{}^2 + C_{DC}R)s + D_{DC-DC}{}^2\right]} \qquad (14)$$

*(ii) Modified wireless power receiver with buck-boost converter*

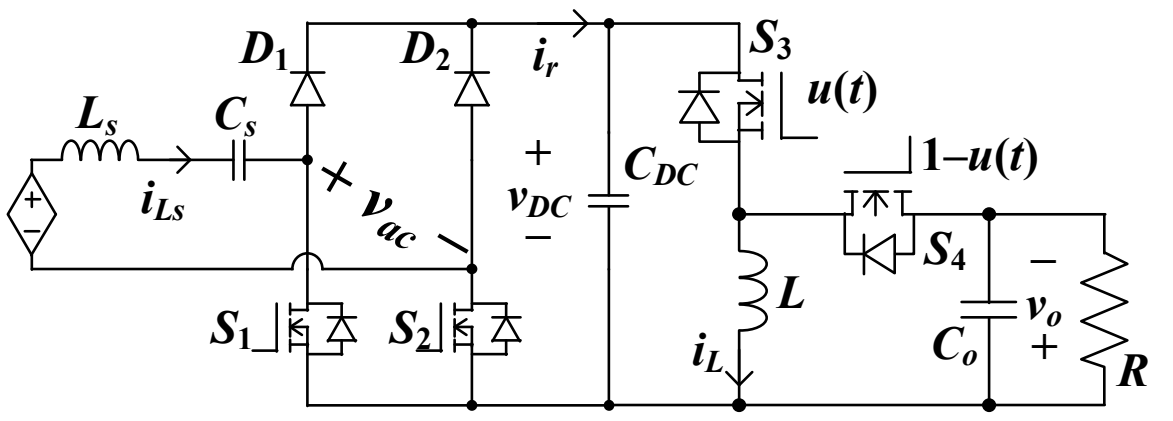

Fig. 6. Modified wireless power receiver with buck-boost converter.

Fig. 6 shows the schematic diagram of the modified wireless power receiver with the buck-boost converter. The state-space averaged model is

$$\begin{bmatrix} \frac{\partial \langle v_{DC} \rangle_T}{\partial t} \\ \frac{\partial \langle i_L \rangle_T}{\partial t} \\ \frac{\partial \langle v_o \rangle_T}{\partial t} \end{bmatrix} = \begin{bmatrix} 0 & -\frac{D_{DC-DC}}{C_{DC}} & 0 \\ \frac{D_{DC-DC}}{L} & 0 & -\frac{1-D_{DC-DC}}{L} \\ 0 & \frac{1-D_{DC-DC}}{C_o} & -\frac{1}{RC_o} \end{bmatrix} \begin{bmatrix} \langle v_{DC}(t) \rangle_T \\ \langle i_L \rangle_T \\ \langle v_o \rangle_T \end{bmatrix} + \begin{bmatrix} \frac{[1-\cos(2\pi \boldsymbol{D})]I_{Ls}}{\pi C_{DC}} \\ 0 \\ 0 \end{bmatrix} \qquad (15)$$

Its control-to-output transfer function is derived as

$$G_{M_{buck-boost}}(s) = \frac{2D_{DC-DC}I_{Ls}R\sin(2\pi D)(1-D_{DC-DC})}{\left[C_oC_{DC}LRs^3 + C_{DC}Ls^2 + (C_oD_{DC-DC}{}^2 + C_{DC}(1-D_{DC-DC})^2)Rs + D_{DC-DC}{}^2\right]} \qquad (16)$$

*(iii) Modified wireless power receiver with boost converter*

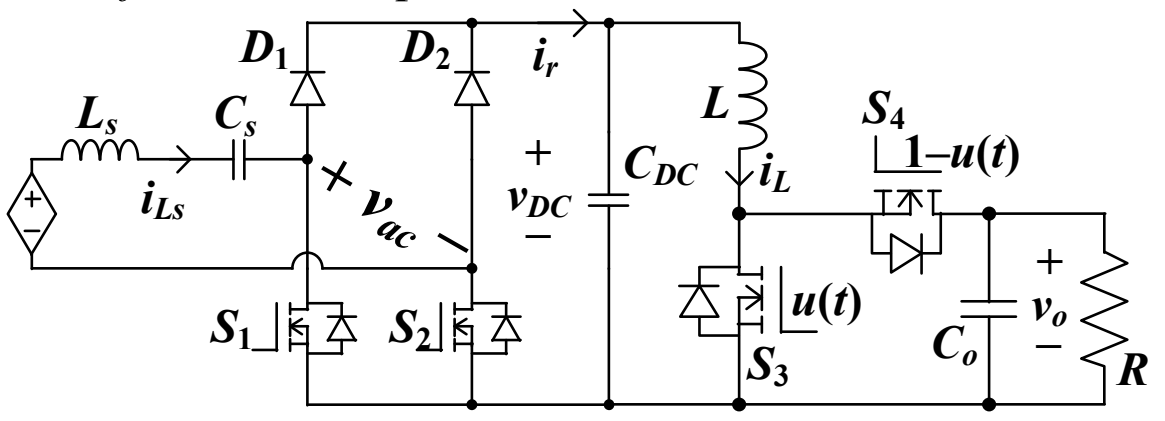

Fig. 7. Modified wireless power receiver with boost converter.

Fig. 7 shows the schematic diagram of the modified wireless power receiver with boost converter. The state-space averaged is

$$\begin{bmatrix} \frac{\partial \langle v_{DC} \rangle_T}{\partial t} \\ \frac{\partial \langle i_L \rangle_T}{\partial t} \\ \frac{\partial \langle v_o \rangle_T}{\partial t} \end{bmatrix} = \begin{bmatrix} 0 & -\frac{1}{C_{DC}} & 0 \\ \frac{1}{L} & 0 & -\frac{1-D_{DC-DC}}{L} \\ 0 & \frac{1-D_{DC-DC}}{C_o} & -\frac{1}{RC_o} \end{bmatrix} \begin{bmatrix} \langle v_{DC}(t) \rangle_T \\ \langle i_L \rangle_T \\ \langle v_o \rangle_T \end{bmatrix} + \begin{bmatrix} \frac{[1-\cos(2\pi \boldsymbol{D})]I_{Ls}}{\pi C_{DC}} \\ 0 \\ 0 \end{bmatrix} \qquad (17)$$

Its control-to-output transfer function is

$$G_{Mboost}(s) = \frac{2I_{Ls}R\sin(2\pi D)(1-D_{DC-DC})}{\left[C_oC_{DC}LRs^3 + C_{DC}Ls^2 + (C_o + C_{DC}(1-D_{DC-DC})^2)Rs + 1\right]} \qquad (18)$$

As compared with the state-space equations of the original wireless power receivers (Eq. (3), (6), (9)) and modified wireless power receivers (Eq. (13), (15), and (17)), the major difference is that the control variable $D_{DC\text{-}DC}$ of the original ones is embedded in the state matrix and coupled with the state variables, while the control variable $D$ of the modified ones is located at the input matrix and decoupled with the state variables. Via decoupling the control variables from the state variables, the RHP zeros can be removed.

To verify the accuracy of the models, the circuit simulation results carried out using PSIM v2020a, and the theoretical results (Eq. (4), (7), (10), (14), (16), and (18)) are plotted altogether as shown in Fig. 8. The parameters used are $I_{Ls}$=1 A, $C_{DC}$= 30 $\mu$F, $L$=77 $\mu$H, $C_o$=40 $\mu$F, $R$=7, $D_{DC\text{-}DC}$=0.5, and $D$=0.51. Fig. 8 (a)-(c) respectively show the theoretical and simulated Bode plots of the transfer functions of the original and modified wireless power receivers with buck converter, buck-boost converter, and boost converter. The simulated and the theoretical results are in close agreement, validating the accuracy of the models.

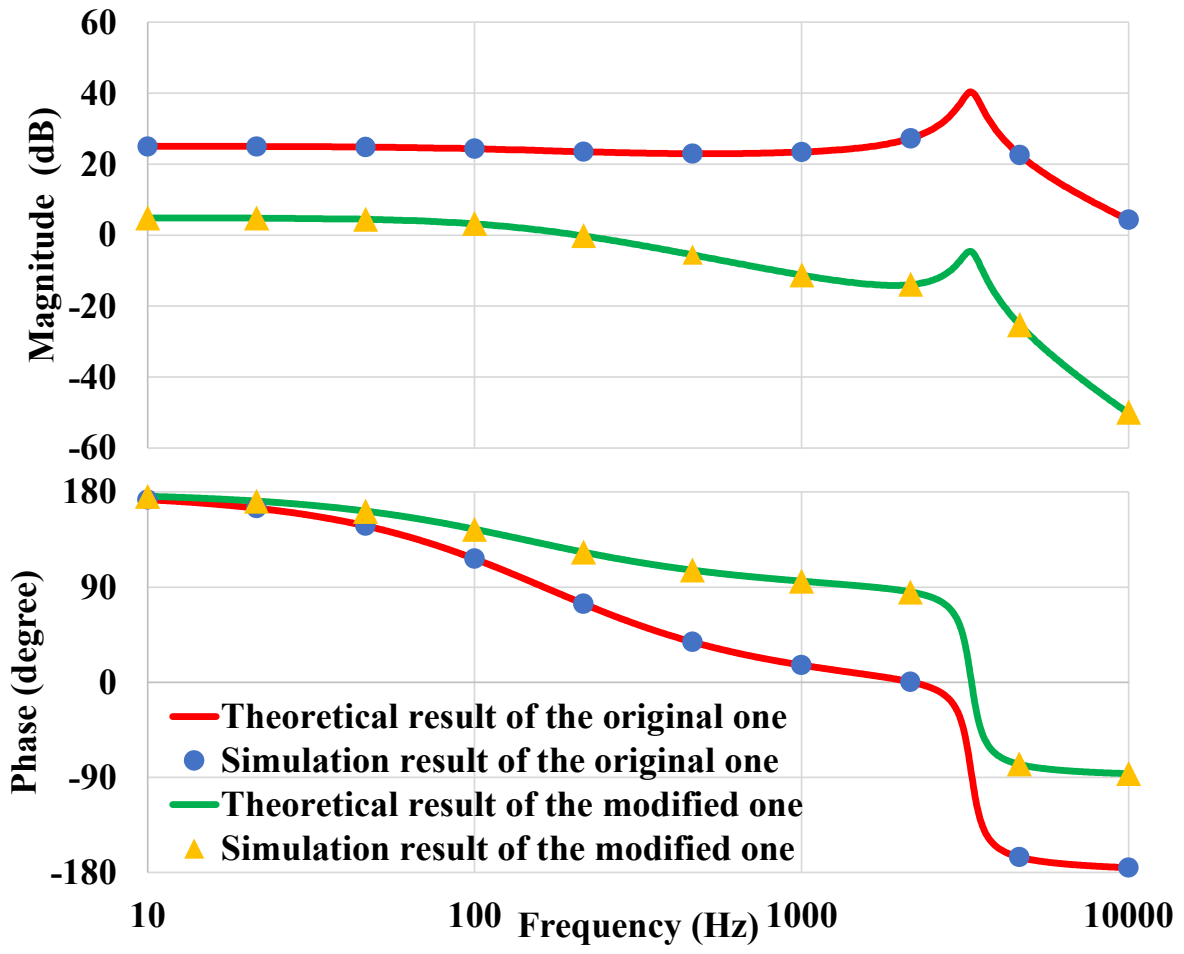

(a) Wireless power receivers with buck converter

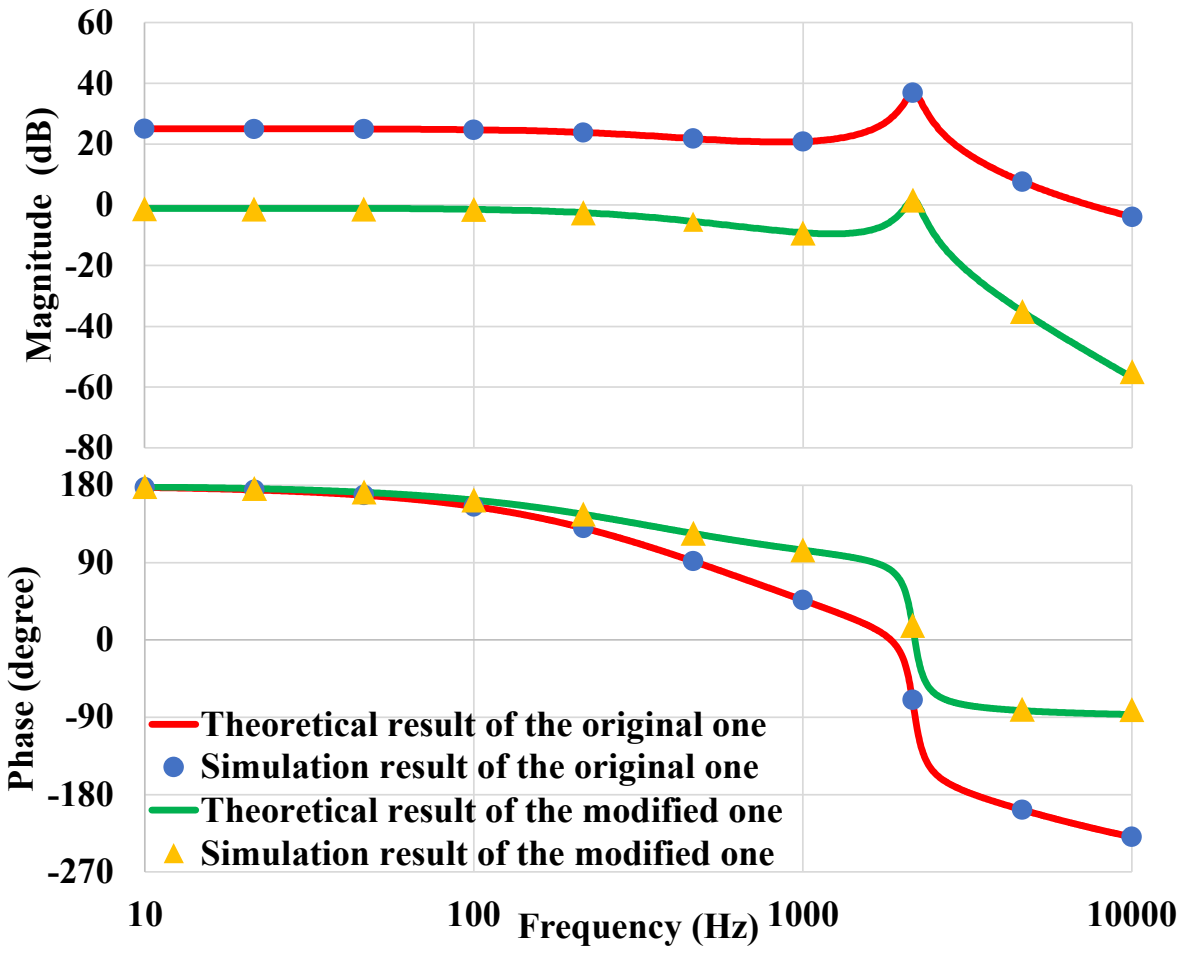

(b) Wireless power receivers with buck-boost converter

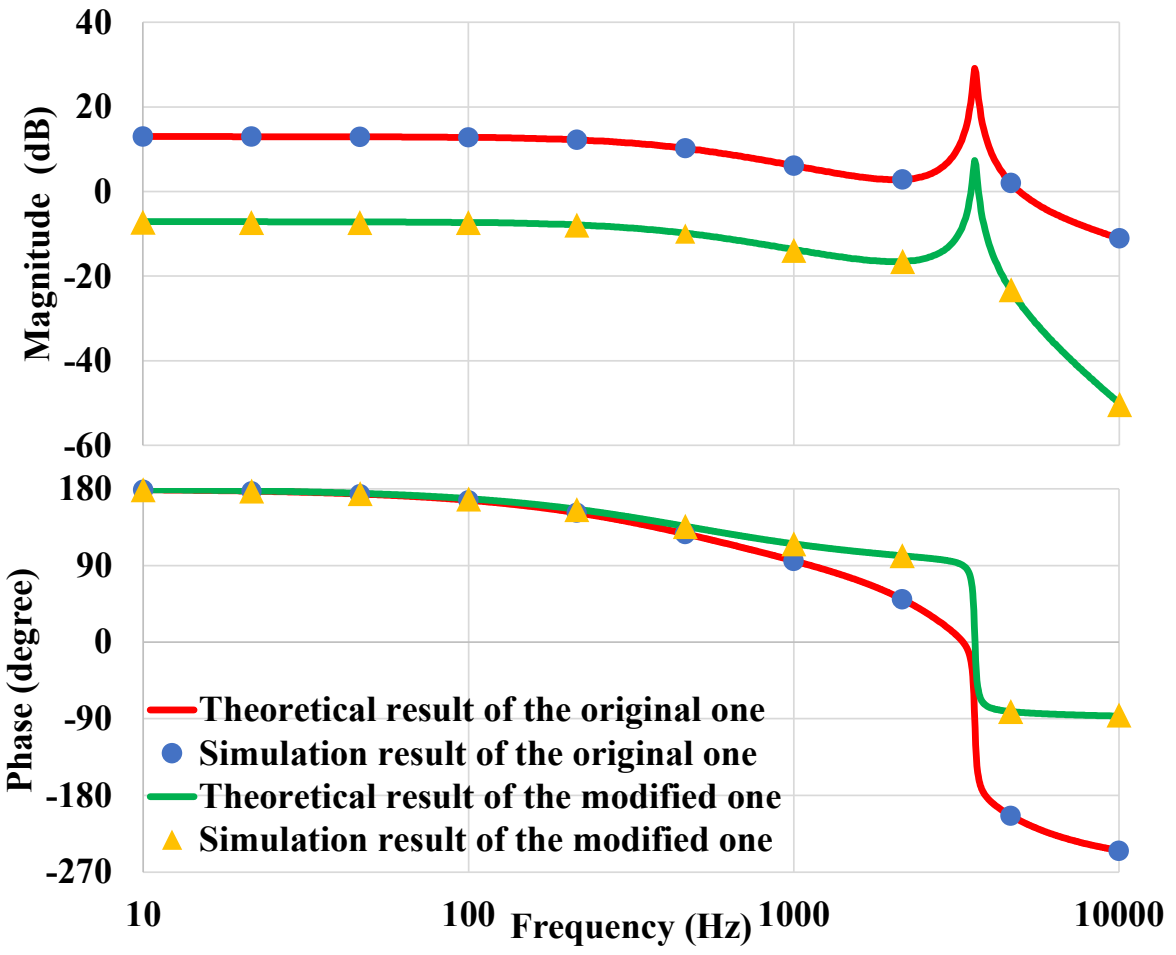

(c) Wireless power receivers with buck-boost converter

Fig. 8. Theoretical and simulated Bode plots the transfer functions of the wireless power receivers.

Moreover, Fig. 9 (a)-(c) show the pole-zero maps of the transfer functions of the original and modified wireless power receivers. The pole-zero maps of the wireless power receivers show that the RHP zeros are eliminated by using modified wireless power receivers. In addition, as compared with the original ones (see Eq. (4), (7), and (10)), the transfer functions of the modified wireless power receiver (see Eq. (14), (16), and (18)) show that the RHP zero are eliminated without affecting the poles.

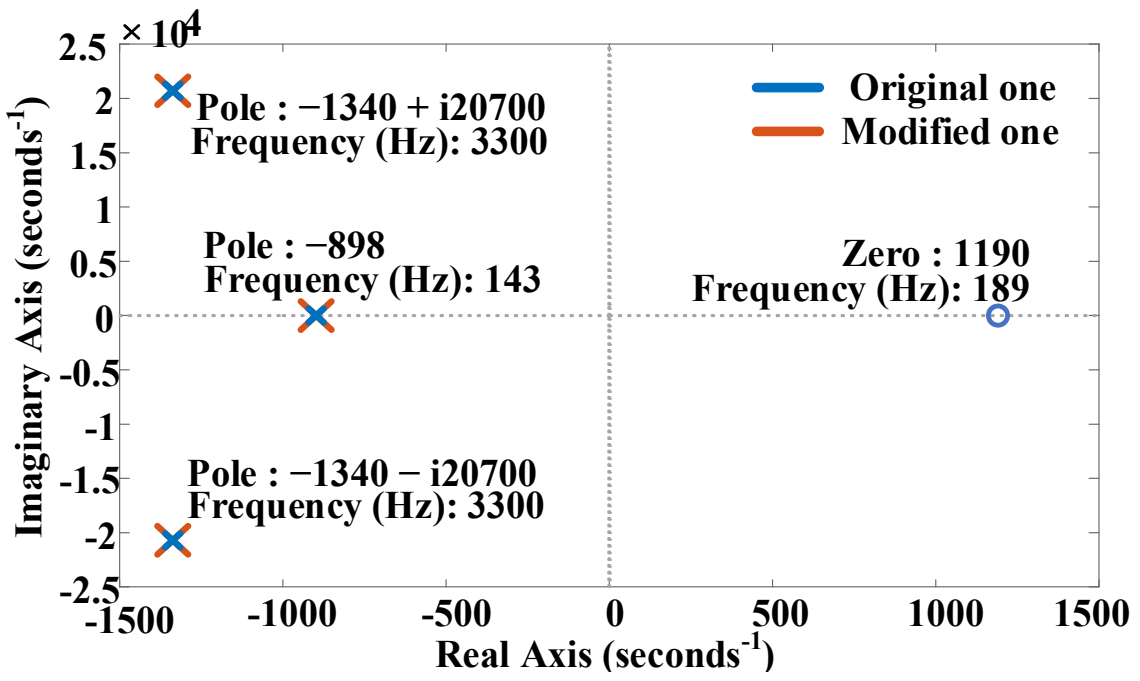

(a) Wireless power receivers with buck converter

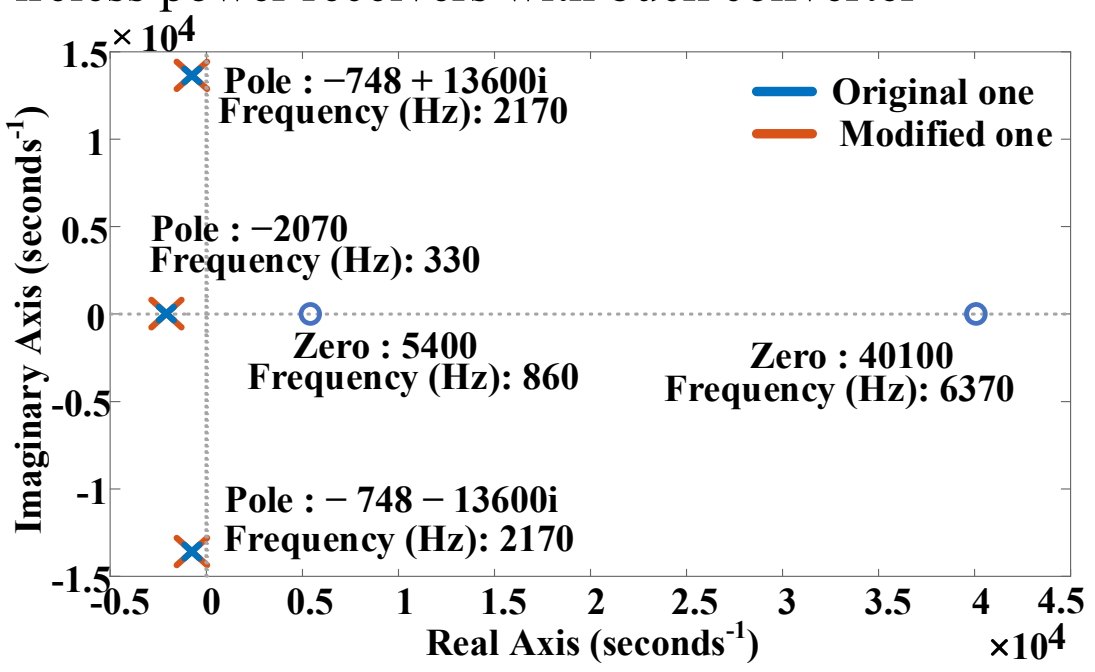

(b) Wireless power receivers with buck-boost converter

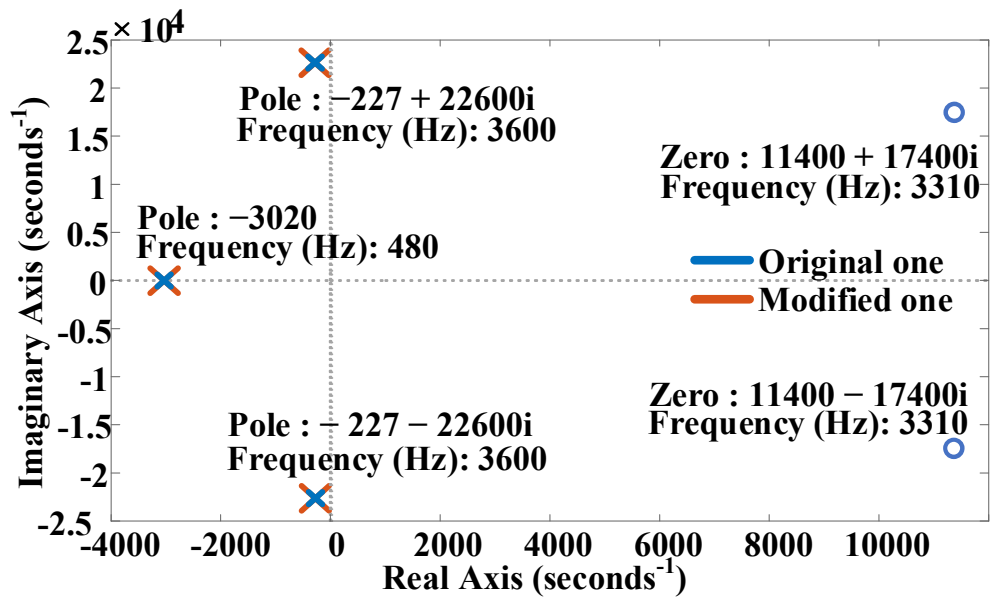

(c) Wireless power receivers with boost converter

Fig. 9. Pole-zero maps of the transfer functions of the wireless power receivers.

## IV. Dynamic Performance Improvement

The stability and dynamic response of wireless power receivers with single-loop feedback control (see Fig. 10) is assessed, of which a proportional-integral (PI) compensator $G_c(s)$ is used to achieve output voltage regulation. The transfer function of the PI compensator is

$$G_c(s) = k_p + \frac{k_i}{s} \tag{19}$$

where $k_p$ and $k_i$ are the proportional gain and integral gain respectively.

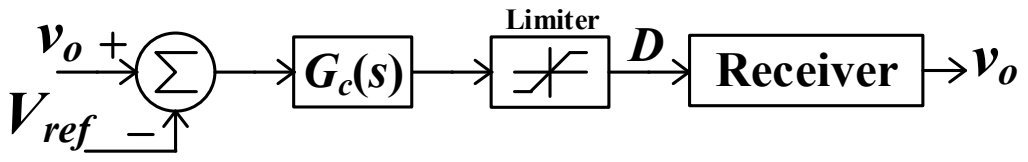

Fig. 10 Block diagram of the closed-loop system.

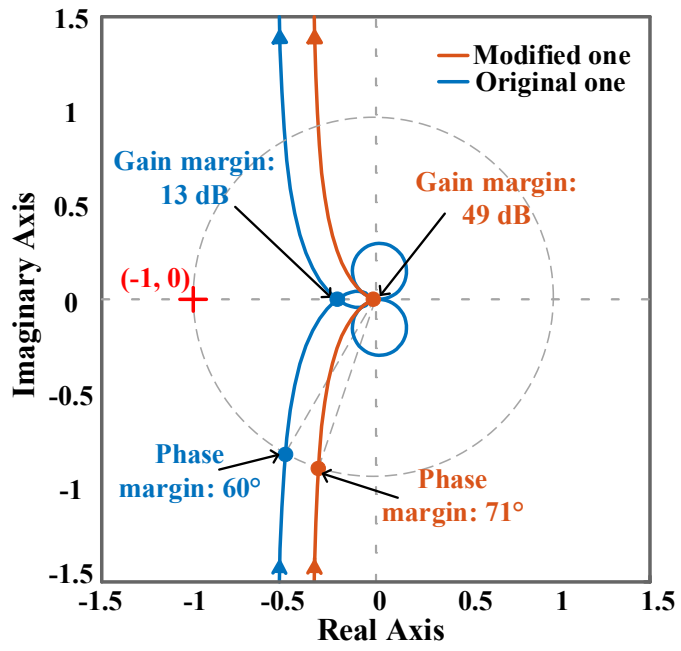

(a) Nyquist plots of the loop gains

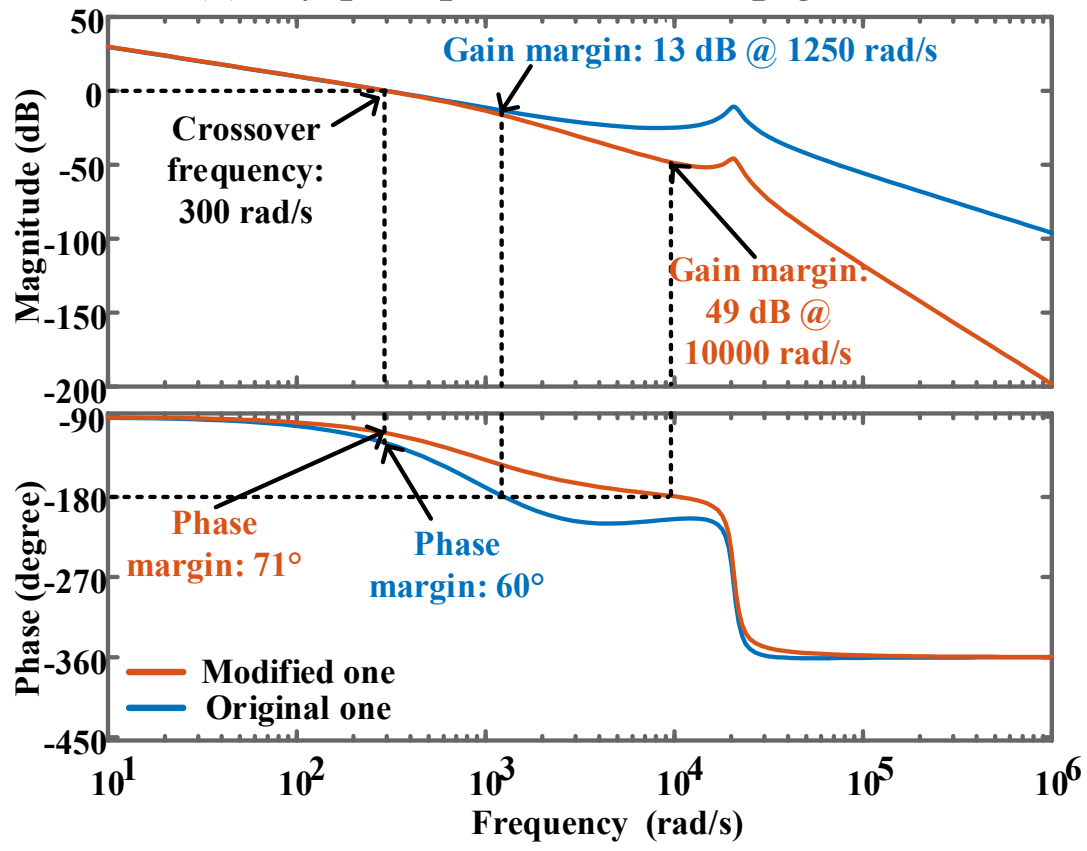

(b) Bode plots of the loop gains

Fig. 11. Bode plots of loop gains of the modified and original wireless power receivers with buck converter.

### A. Closed-Loop Stability Assessment

To have a fair assessment of the closed-loop stability, the compensators are designed such that the crossing frequency of loop gains of wireless power receivers are located at the same frequency. To ensure sufficient stability margin for both the original and modified wireless power receivers, the crossing frequency is chosen at around 1/4 of the RHP zero (1190 rad/s, see Fig. 9 (a)) of the buck converter. Consequently, the crossing frequency is selected at 300 rad/s.

Fig. 11 shows the characteristics of the loop gains of the original and modified wireless power receivers with buck converter. The parameters of the compensator of the original receiver are $k_p$=0.0027284 and $k_i$=17.1836, and that of the modified receiver are $k_p$=0 and $k_i$=179753. Fig. 11(a) shows no encirclement of (−1, 0), suggesting that both receivers are stable. Fig. 11 (b) shows that the modified one features a phase margin of 71° at 300 rad/s, and a gain margin of 49 dB at 10000 rad/s, while the original wireless power receiver features a phase margin of 60° at 300 rad/s and a gain margin of 13 dB at 1250 rad/s.

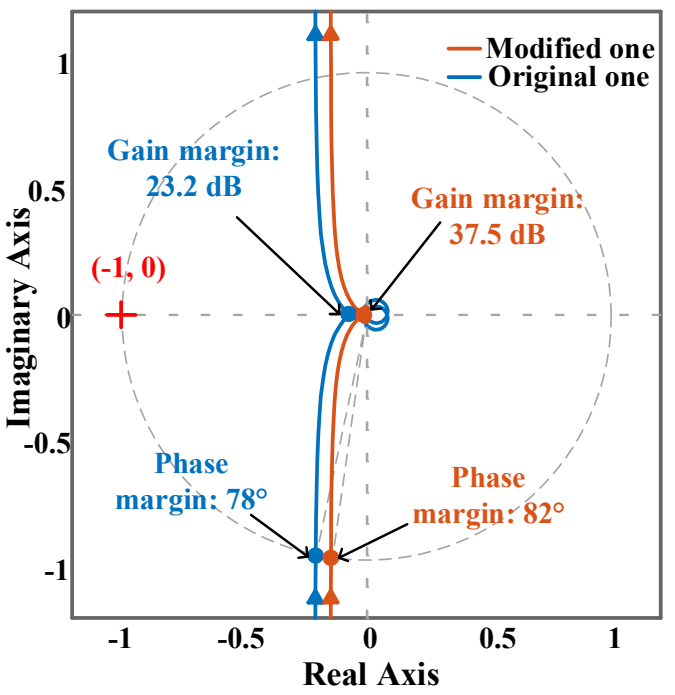

(a) Nyquist plots of the loop gains

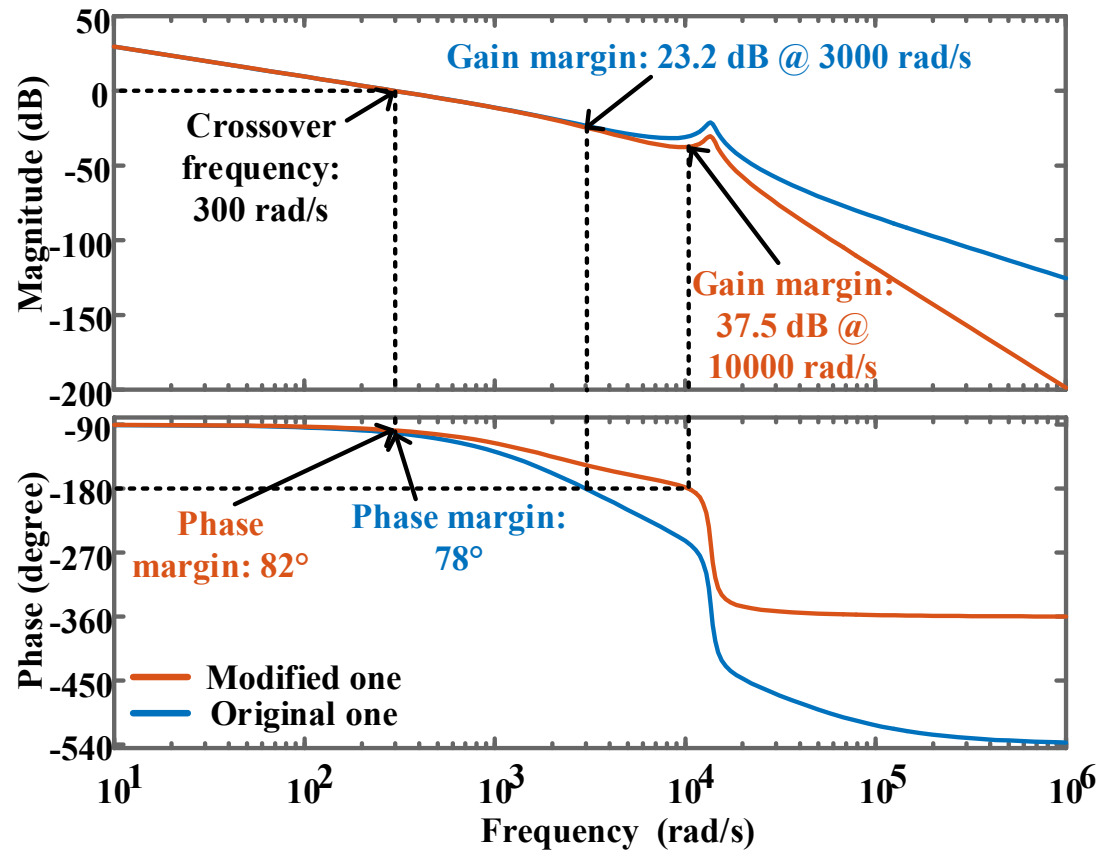

(b) Bode plots of the loop gains

Fig. 12. Bode plots of loop gains of the modified and original wireless power receivers with buck-boost converter.

Fig. 12 shows the characteristics of the loop gains of the original and modified wireless power receivers with buck-boost converter. The parameters of the compensator are $k_p$=0 and $k_i$=16.97 and $k_p$=0 (original receiver) and $k_i$= 344427 (modified receiver). Fig. 12 (a) shows no encirclement of (−1, 0), suggesting that both receivers are stable. Fig. 12 (b) shows that the modified one features a phase margin of 82° at 300 rad/s, and a gain margin of 37.5 dB at 10000 rad/s, while the original wireless power receiver features a phase margin of 78° at 300 rad/s and a gain margin of 23.2 dB at 3000 rad/s.

Fig. 13 shows the characteristics of the loop gains of the original and modified wireless power receivers with boost converter. The parameters of the compensator are $k_p$=0 and $k_i$=67.64 (original receiver) and $k_p$= 0 and $k_i$= 685335 (modified receiver). Fig. 13 (a) shows no encirclement of (−1, 0), suggesting that both receivers are stable. Fig. 13 (b) shows that the modified one features a phase margin of 84° at 300 rad/s, and a gain margin of 37.5 dB at 20800 rad/s, while the original wireless power receiver features a phase margin of 83° at 300 rad/s and a gain margin of 34.9 dB at 7050 rad/s.

In general, both the original and modified wireless power receivers are stable. However, the modified ones show larger phase margins and gain margins, suggesting improved stability margins.

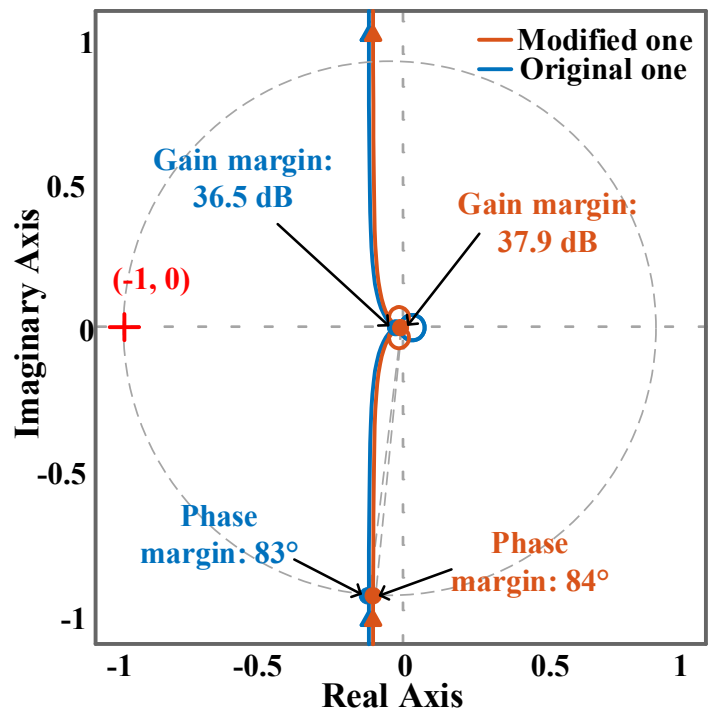

(a) Nyquist plots of the loop gains

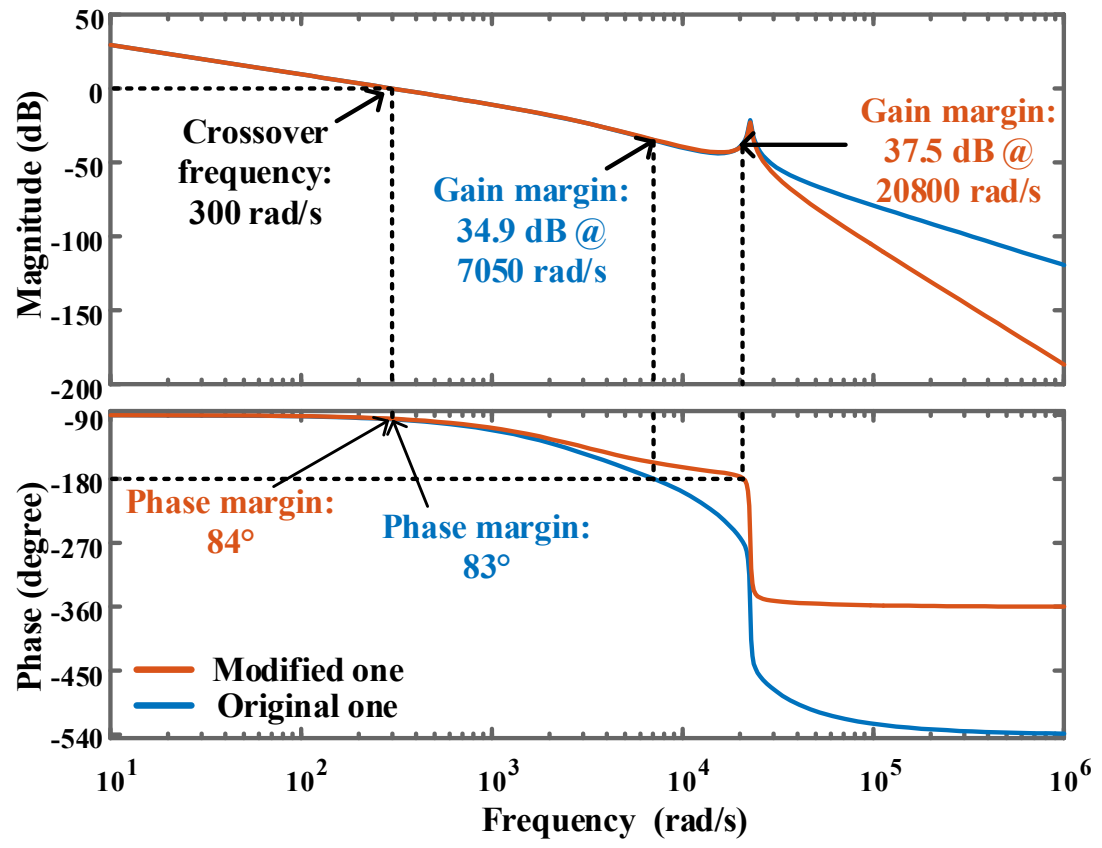

(b) Bode plots of the loop gains

Fig. 13. Bode plots of loop gains of the wireless power receivers.

## *B. Dynamic Response Assessment*

For fair assessment of dynamic responses, the compensator is designed such that the set of phase margin and gain margin of wireless power receivers are identical.

Fig. 14(a) shows the Bode plots of the loop gains of the wireless power receivers with buck converter, where both receivers share identical gain margin of 20 dB and phase margins of 76.8°. The parameters of the compensator are $k_p$=0 and $k_i$=6.64 (original receiver) and $k_p$=0.0732 and $k_i$=130.25 (modified receiver). However, the modified one achieves a higher crossover frequency of 480 rad/s, while the original one achieves a relatively low crossover frequency of 118 rad/s. Fig. 14(b) shows the time-domain response of voltage reference step change from 8 V to 8.8 V. The modified one reaches the new reference value faster than the original one. These results validate that the modified wireless power receiver can improve the dynamic response.

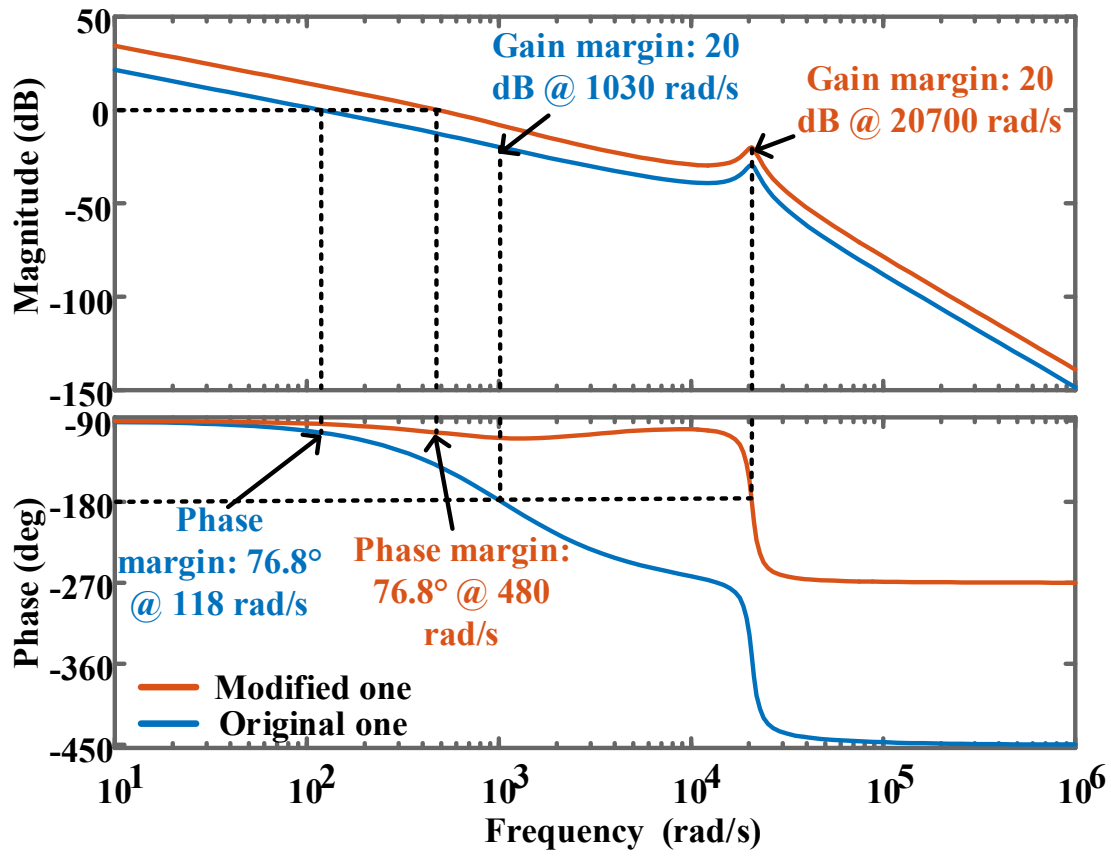

(a) Bode plots of the loop gains

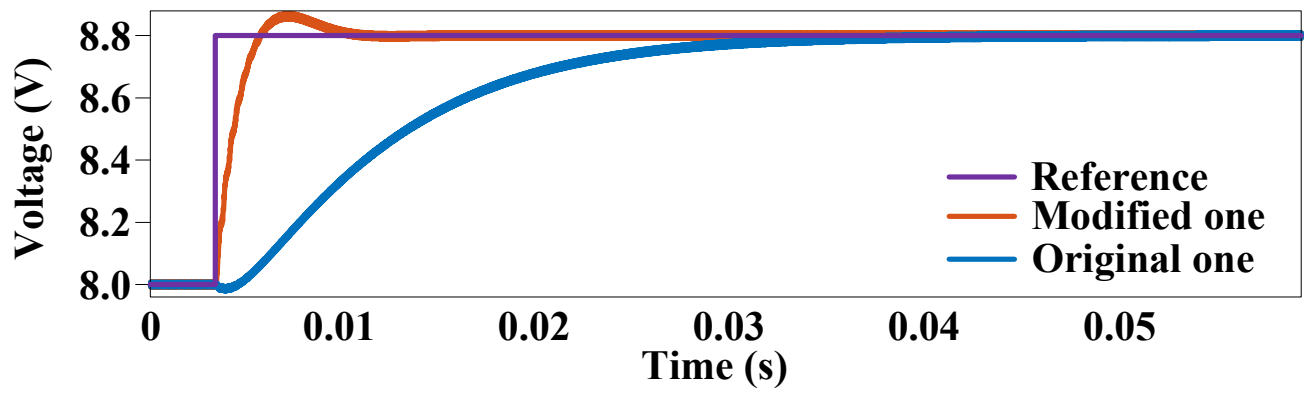

(b) Time-domain waveforms of reference tracking

(c)

Fig. 14. Frequency-domain and time-domain responses of the original and modified wireless power receivers with buck converter.

Fig. 15(a) shows the Bode plots of the loop gains of the wireless power receivers with buck-boost converter, where both receivers are compensated such that they have identical gain margins of 20 dB and phase margins of 72.9°. The parameters of the compensator are $k_p$=0 and $k_i$=24.53 (original receiver) and $k_p$=0.0167 and $k_i$=228.36 (modified receiver). However, the modified one achieves a higher crossover frequency of 751 rad/s, while the original one achieves a relatively low crossover frequency of 430 rad/s. Fig. 15(b) shows the time-domain response of voltage reference step change from 4 V to 4.4 V. The modified one reaches the new reference value faster than the original one. These results validate that the modified wireless power receiver can improve the dynamic response.

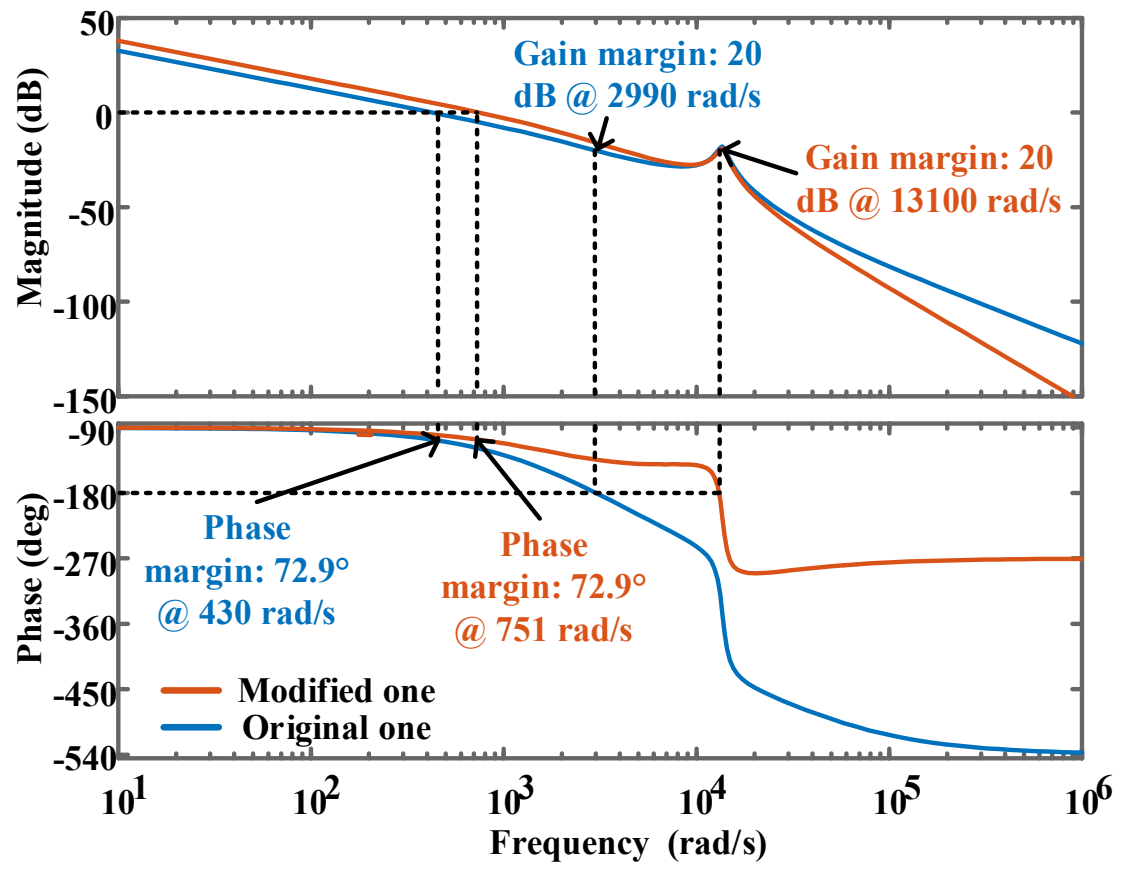

(a) Bode plots of the loop gains

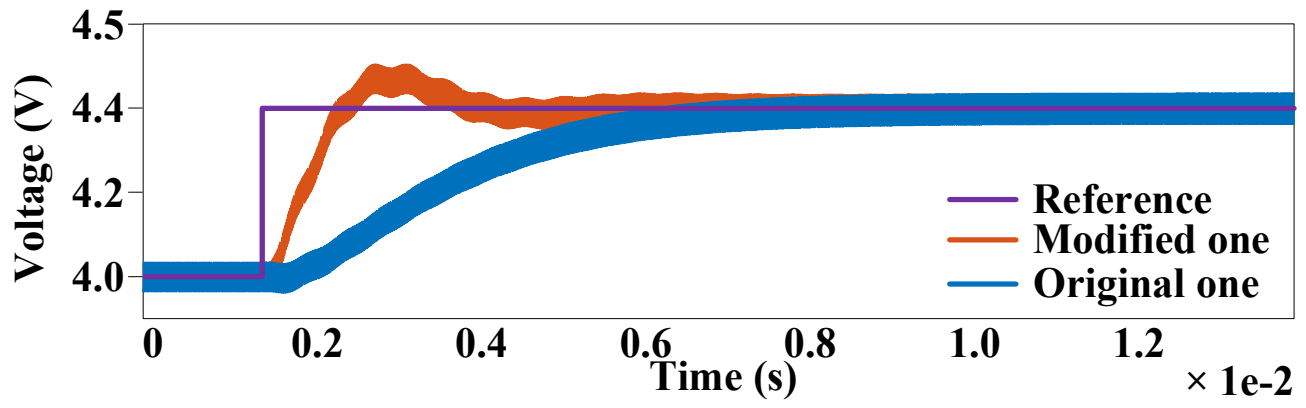

(b) Time-domain waveforms of reference tracking

Fig. 15. Frequency-domain and time-domain responses of the original and modified wireless power receivers with buck-boost converter.

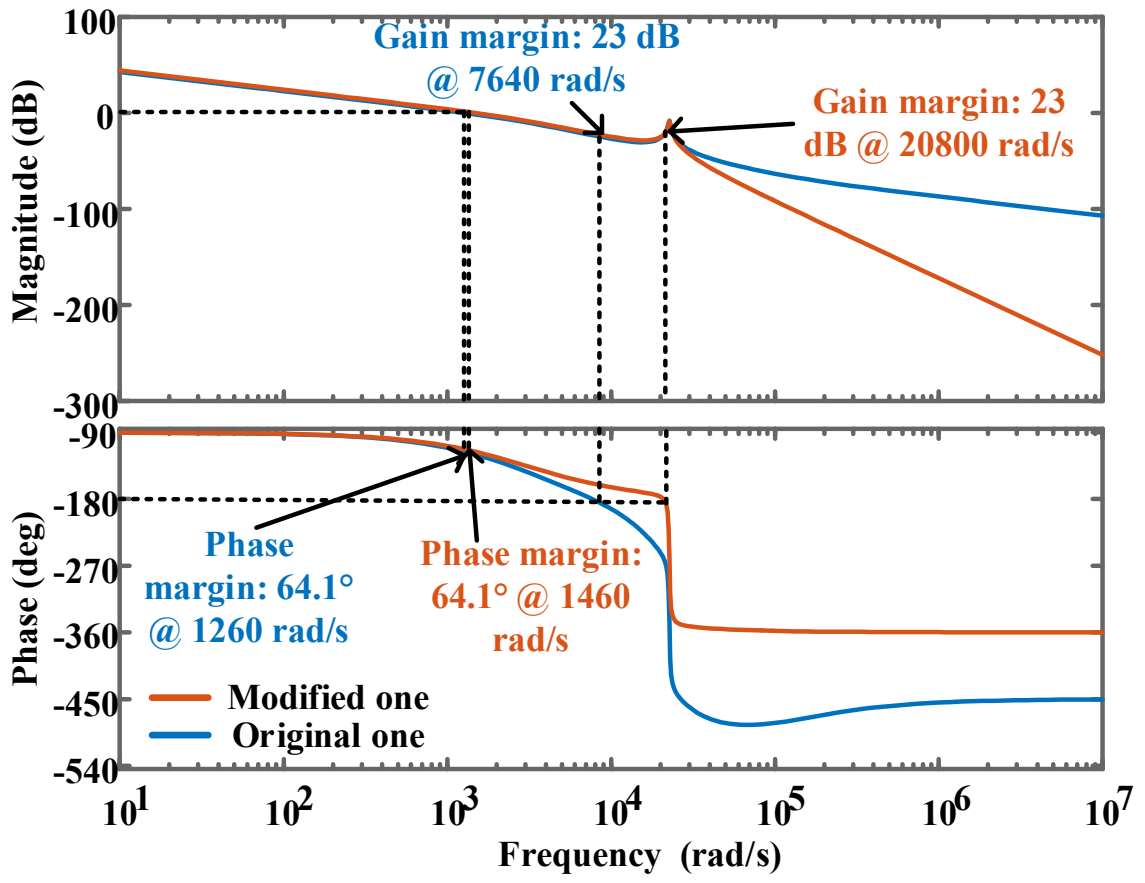

(a) Bode plots of the loop gains

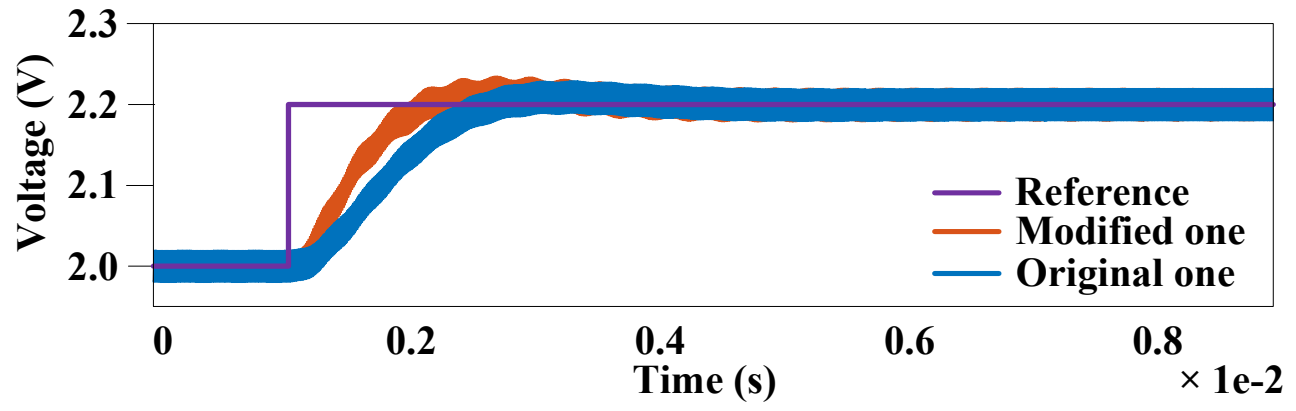

(b) Time-domain waveforms of reference tracking

Fig. 16. Frequency-domain and time-domain responses of the original and modified wireless power receivers with boost converter.

Fig. 16(a) shows the Bode plots of the loop gains of the wireless power receivers with boost converter, where both receivers are compensated such that they have identical gain margins of 23 dB and phase margins of 64.1°. The parameters of the compensator are $k_p$=0.002777 and $k_i$=305.88 (original receiver) and $k_p$=0 and $k_i$=745.744 (modified receiver). However, the modified one achieves a higher crossover frequency of 1460 rad/s, while the original one achieves a relatively low crossover frequency of 1260 rad/s. Fig. 16(b) shows the time-domain response of voltage reference step change from 2 V to 2.2 V. The modified one reaches the new reference value slightly faster than the original one. These results show that the modified wireless power receiver can improve the dynamic response.

## V. Experimental Verification

TABLE I. List of Components

| Component / Parameters | Value / Part Number |
|---|---|
| $L_s$ | 103 $\mu$H Hand-made |
| $C_s$ | 24 nF<br>B32672L1123J000×2 |
| $C_{DC}$ | 30 $\mu$F<br>CL32B106KBJNNNE×3 |
| $L$ | 77 $\mu$H Hand-made |
| $C_o$ | 40 $\mu$F<br>CL32B106KBJNNNE×4 |
| Gate Driver | ADuM3223 |
| MOSFET ($S_1$, $S_2$) | IPP055N03LGXKSA1 |
| MOSFET ($S_3$, $S_4$) | IRLU3714ZPBF |
| Diode ($D_1$-$D_4$) | IPP055N03LGXKSA1 |
| MCU | LAUNCHXL-F28379D |

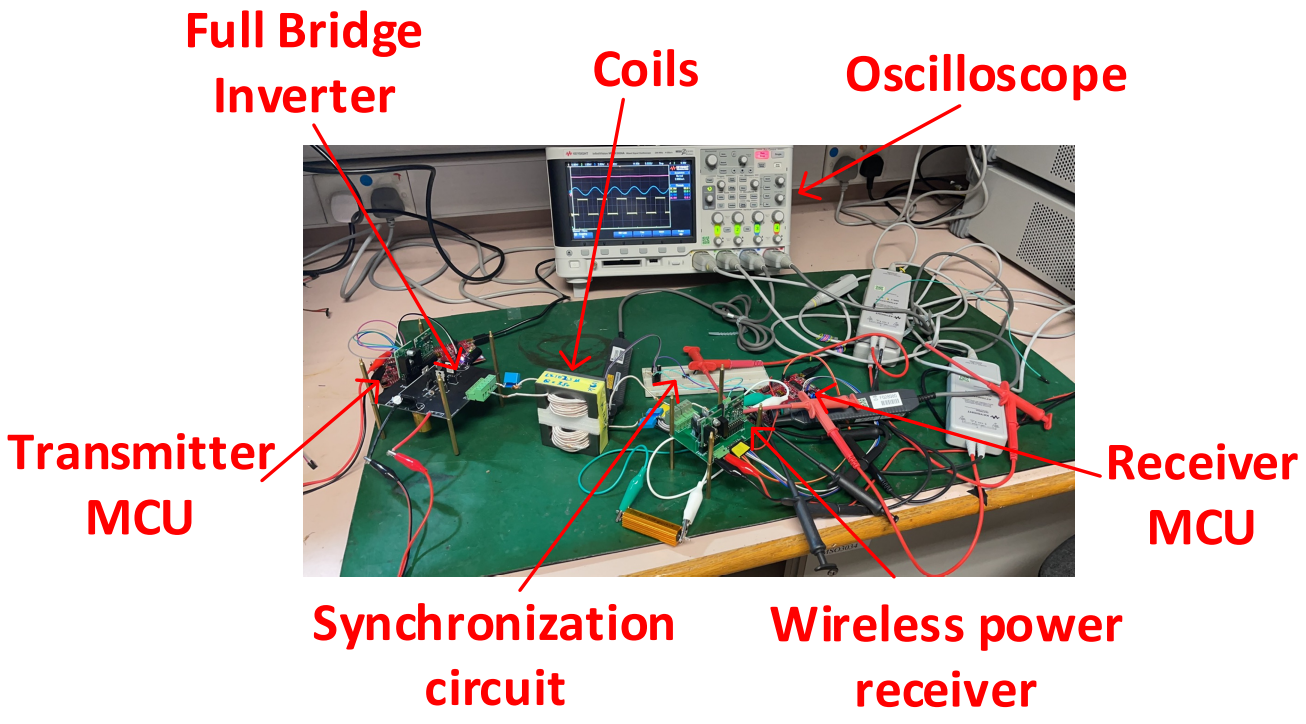

Fig. 17. Photograph of the prototype and the experimental setup.

Fig. 17 shows the experimental setup, of which an MSXO3024A oscilloscope, N2873 passive voltage probes, N2790A differential voltage probes, and 1147B current probes are used for the measurement. The modified and original wireless power receivers with buck converter are implemented and verified in the experiment. The components and parameters of the prototype are shown in Table I.

### *A. Open-Loop Verification*

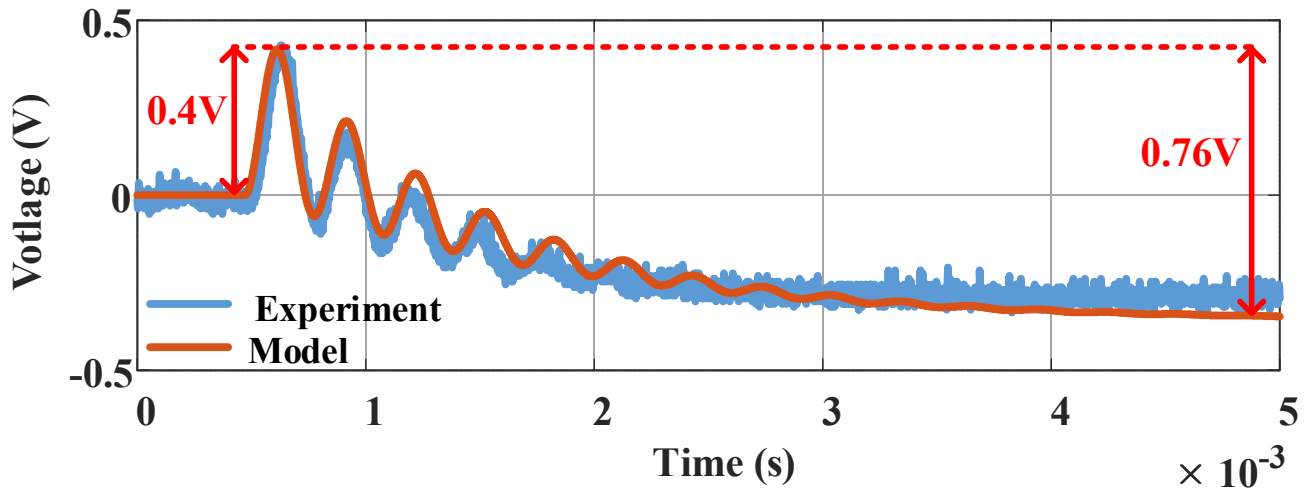

(a) Wireless power receiver with buck converter

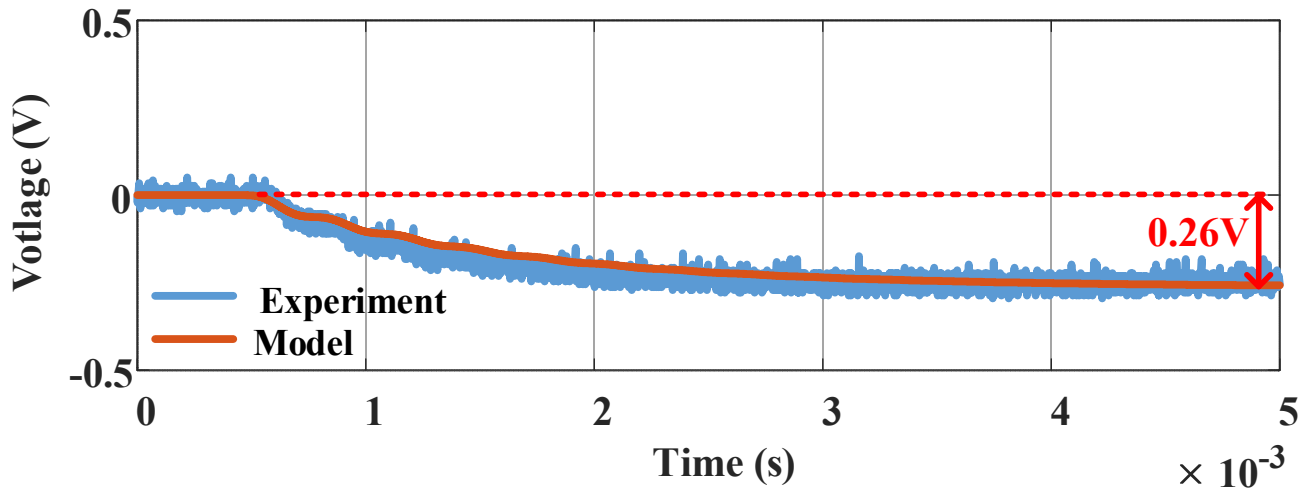

(b) Modified wireless power receiver with buck converter

Fig. 18. Output voltage waveforms of step responses of wireless power receivers.

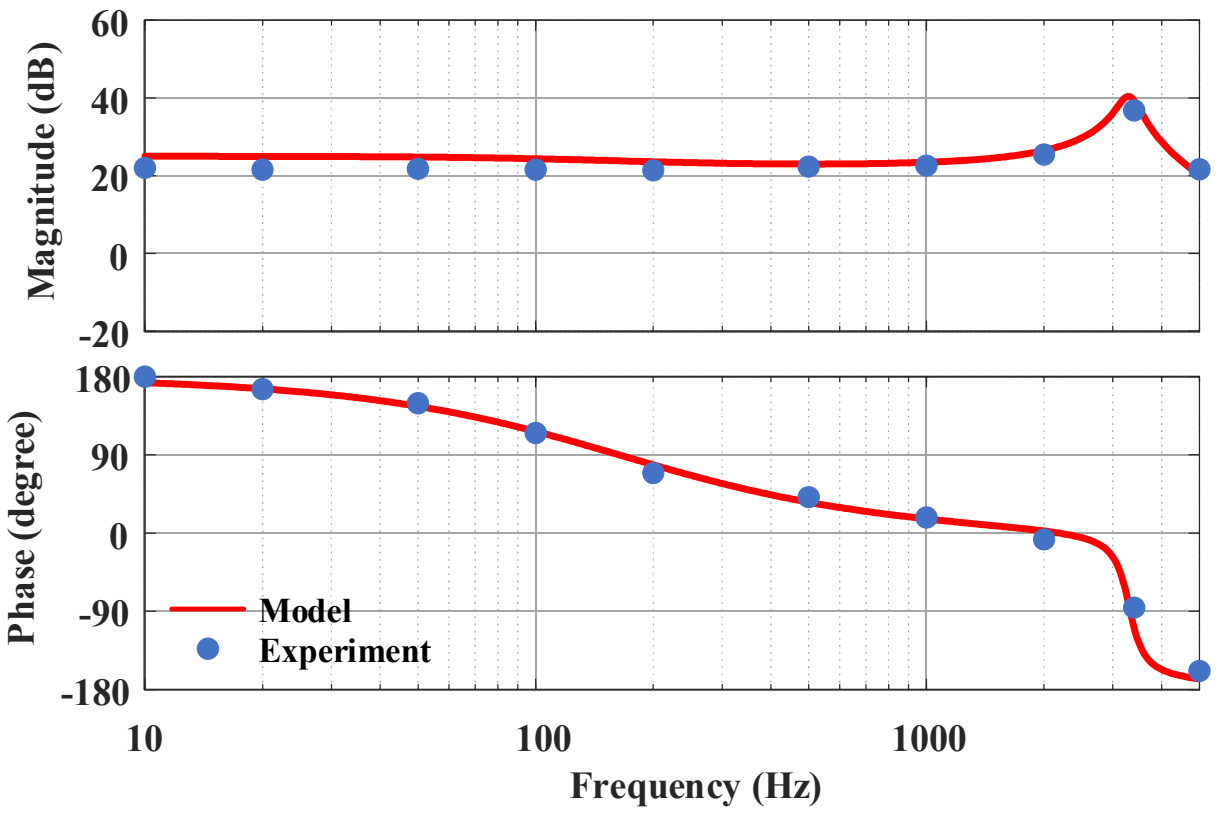

(a) Wireless power receiver with buck converter

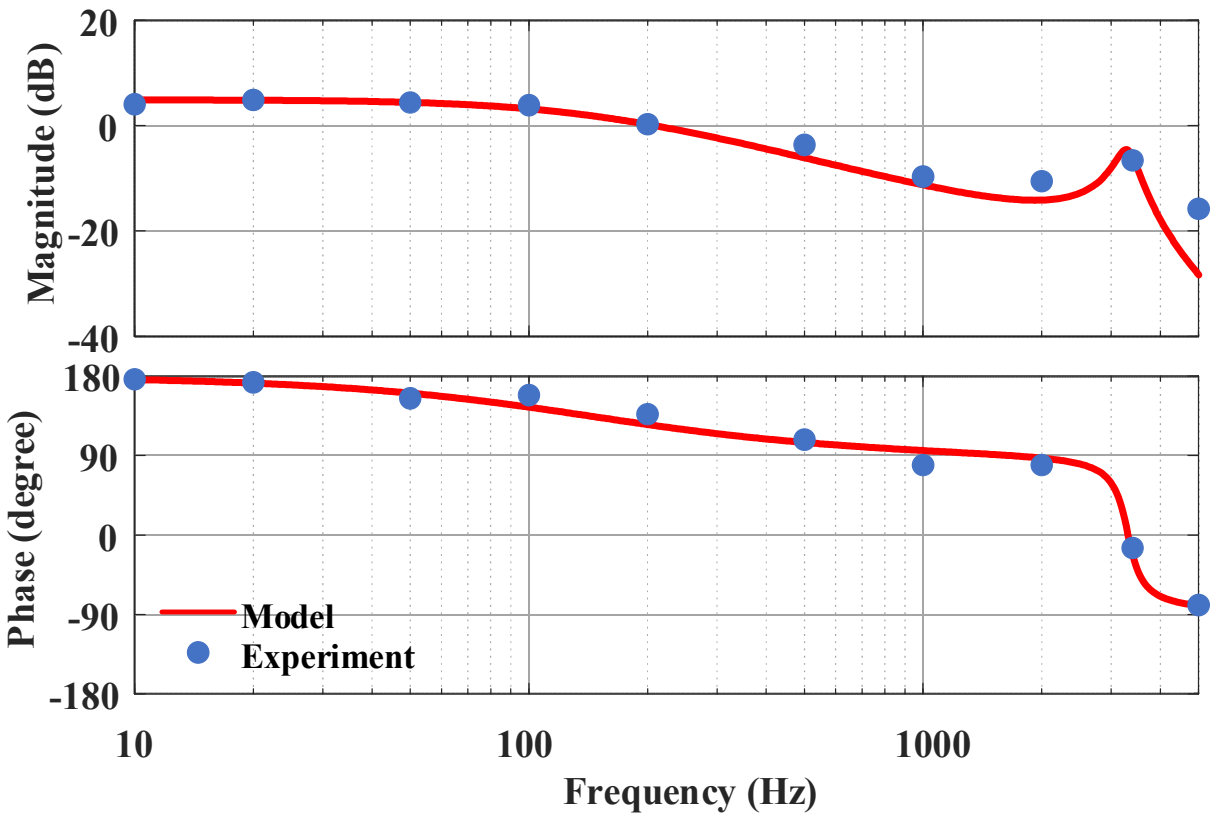

(b) Modified wireless power receiver with buck converter

Fig. 19. Theoretical and experimental Bode plots of wireless power receivers.

The open-loop verifications are performed under these conditions: $I_L$=1 A, $R$=7 Ω, $D_{DC\text{-}DC}$=0.5, and $D$=0.53. Fig. 18(a) shows the output voltage waveform $v_o$ of the wireless power receivers with buck converter in response to the step change of $D_{DC\text{-}DC}$ from 0.5 to 0.52. The output voltage of the wireless power receiver with buck converter shows an inverse response (see Fig. 18(a)), which initially raises by 0.4 V, then drops by 0.76 V, and eventually reaches the new steady-state value. This inverse response behavior signifies the presence of the RHP zero [12], [16]. Fig. 18(b) shows the output voltage waveforms $v_o$ of the modified wireless power receiver with buck converter in response to the step change of $D$ from 0.53 to 0.58. The output voltage does not present an inverse response (see Fig. 18 (b)), which drops by 0.26 V, and eventually reaches the new steady-state value. This monotonical response qualitatively verifies the elimination of the RHP zero. Moreover, as shown in Fig. 18(a) and (b), the experimental results (in blue) are very close to the theoretical results (in red), which are simulated using the small-signal models given in (4) and (14). These results not only validate the accuracy of the models, but further verify the elimination of the RHP zero.

The experimental and theoretical Bode plots of the original and modified wireless power receiver system with buck converter are shown in Fig. 19(a) and (b), respectively. Fixed-frequency sinusoidal small-signal perturbations are injected into the control variables and the corresponding small-signal responses are measured via oscilloscope. Through repeating the measurement at different frequencies, the experimental Bode plots are obtained [12]. The experimental and theoretical results show close agreement, verifying the accuracy of the models. Moreover, comparing the Bode phase plots of Fig. 19(a) and (b), the phase drop of the modified one (from 180° to −90°) is less than that of the original one (from 180° to −180°). The reduction of 90° verifies the elimination of RHP zero.

*B. Closed-Loop Verification*

For fair comparison, both receivers are compensated such that the set of gain and phase margin of both systems are 20 dB and 76.8° (see Fig. 11). The implemented parameters of the compensator are $k_p$=0 and $k_i$=6.6 (original receiver) and $k_p$=0.07 and $k_i$=130 (modified receiver). Fig. 20 shows the dynamic responses of the step change of the reference voltage from 8 V to 8.8 V. Both receivers can ramp up and reach the new steady-state output voltage. The original receiver features a settling time of around 25 ms without overshoot, while the modified one features a settling time of around 5 ms with a slight overshoot of 0.2 V. In general, the modified one shows 5-time faster dynamic response in terms of reference tracking.

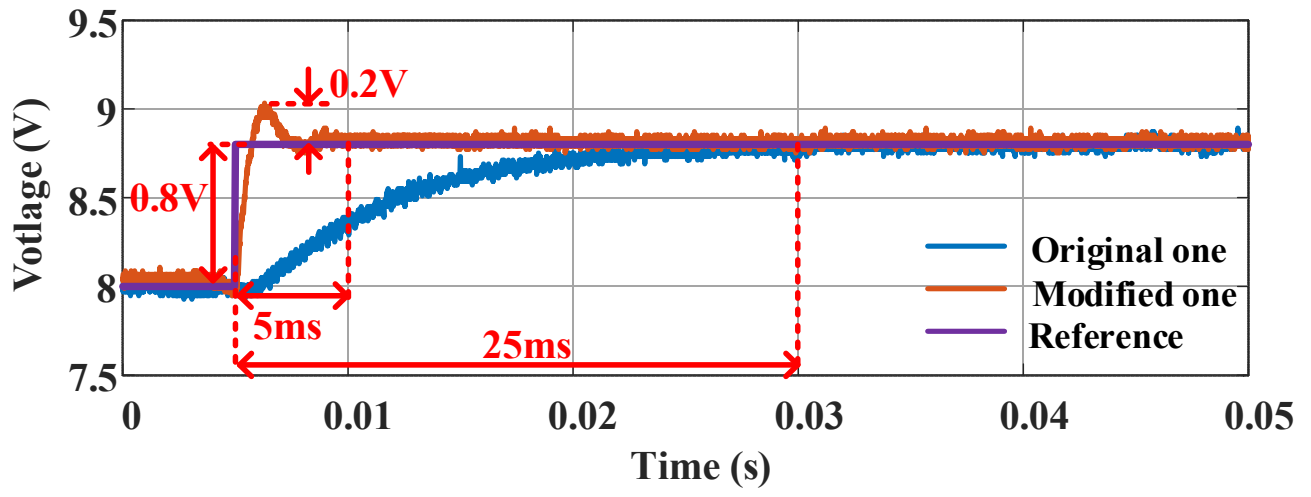

Fig.20. Dynamic responses of the step change of the reference voltage.

Fig. 21 shows the dynamic responses of load step changes from 8.6 Ω to 7 Ω. Both receivers can retain regulated output voltage at 8.8 V. However, the modified one features a settling time of around 4 ms with an undershoot of 0.4 V, while, in contrast, the original one features a settling time of around 25 ms with an undershoot of 1.6 V. The comparative study shows that the modified one features 5-time faster dynamic response and 4-time smaller undershoot voltage against load disturbance.

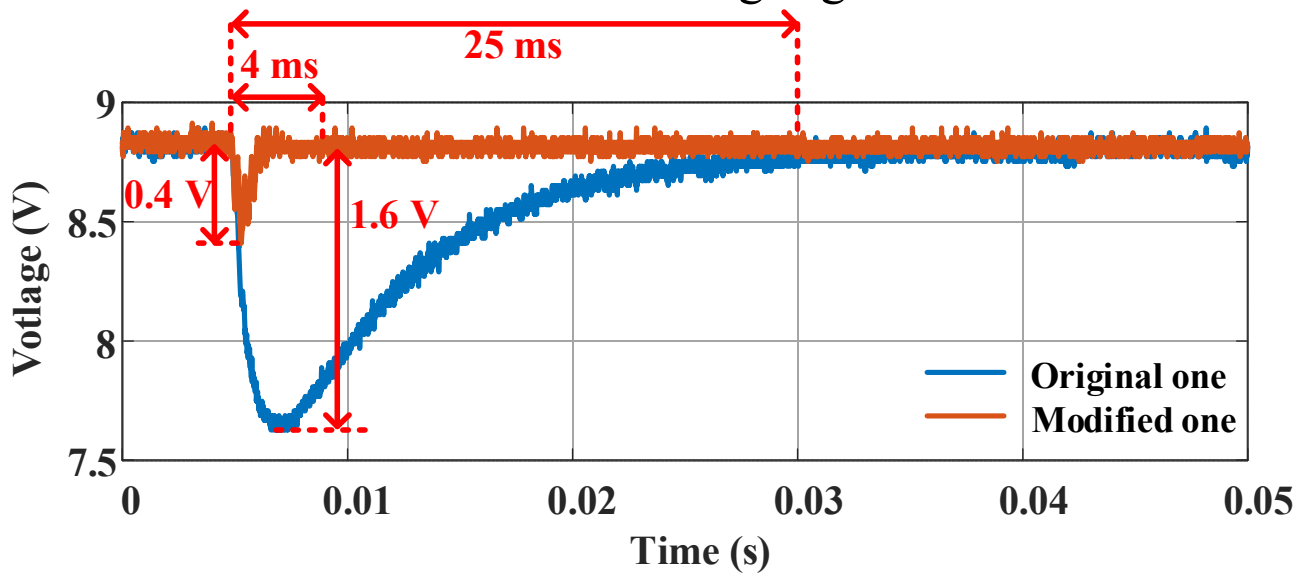

(a) Output voltage

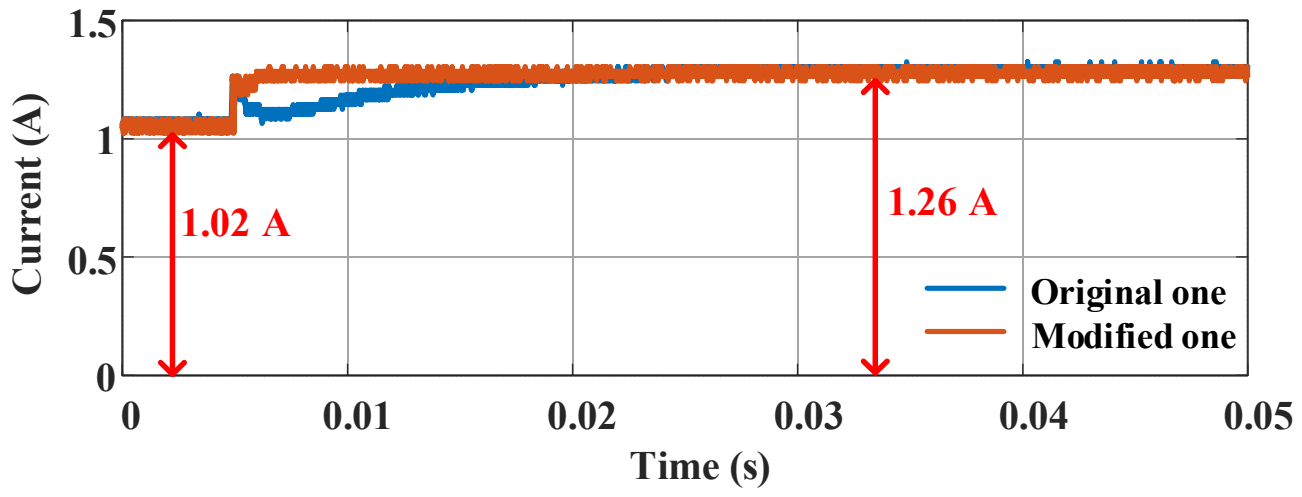

(b) Output current

Fig.21. Dynamic responses of load disturbances under output voltage regulation.

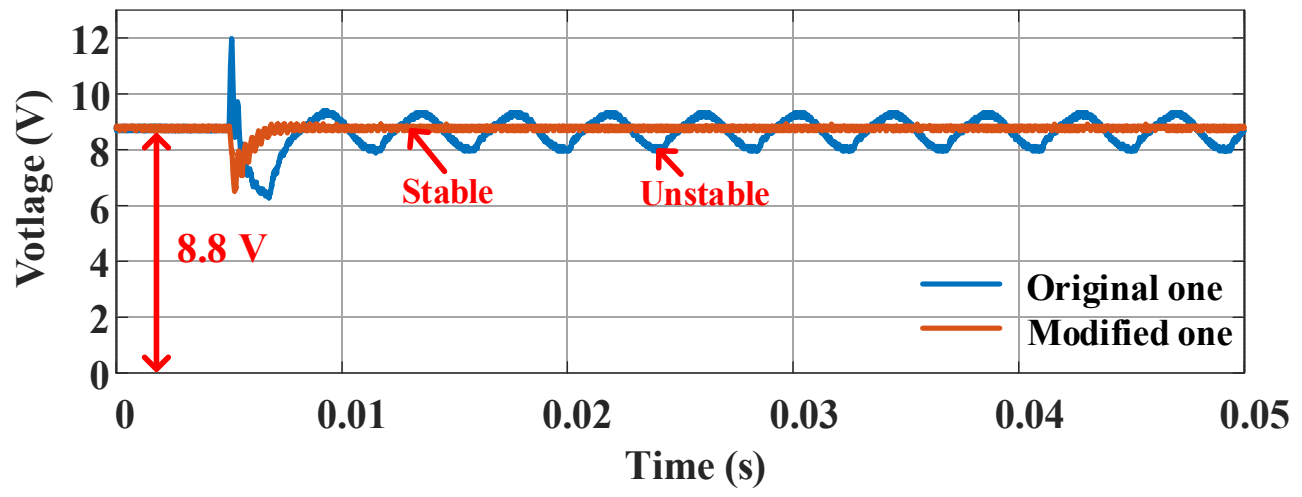

Fig. 22. Dynamic responses in response to the step change of crossover frequency.

To assess the stability margin, Fig. 22 gives the dynamic responses in response to the step change of crossover frequency, after which the crossover frequency of both receiver systems are increased to 1030 rad/s. Correspondingly, the parameters of the compensator of the original receiver are changed from $k_p$= 0 and $k_i$=6.6 to $k_p$= 0 and $k_i$=66, while the parameters of the compensator of the modified receiver are changed from $k_p$= 0.07 and $k_i$= 130 to $k_p$= 0.175 and $k_i$= 325. The output voltage of the original wireless power receiver with buck converter oscillates and becomes unstable, while the output voltage of the modified one remains constant and stable. This test verifies that the modified solution features a larger stability margin.

## VI. Comparison and Discussion

A comparative study of different control methods of the wireless power receiver system is shown in Table II. Reference [1] presents a dual-side control architecture for output voltage regulation where the transmitter provides the coarse adjustment, and the receiver attains fine output regulation. However, due to the latency of wireless communication (from a few milliseconds to hundreds of milliseconds [24]), the dynamic response is slow and stability margin is narrow. [12] presents a dual-loop control method for output voltage regulation. Owing to the poles cancellation and system order reduction property, the dynamic response can be slightly improved but remains limited by the RHP zero. [18] and [19] present nonlinear controllers for DC-link voltage and output voltage regulation, respectively. Large-signal stability can be expected, but they suffer from high design complexity and computation burden. Comparing with these methods, this work presents a general solution for output voltage regulation. Via elimination of the RHP zeros, the closed-loop bandwidth, phase margin and gain margin can be significantly improved, thereby enabling a fast dynamic response and wide stability margin.

## VII. Conclusions

In this paper, the RHP zeros of wireless power receivers with DC-DC converter are investigated. By modifying the power receiver through the use of an active diode rectifier that can control the AC input current, the inherent RHP zeros of the wireless power receivers can be eliminated to achieve good stability and dynamic response of the wireless power receivers. Our experimental studies show that the modified solution can achieve faster dynamic responses in terms of reference tracking and load disturbances under output voltage regulation. In addition, the modified receiver features a larger stability margin under the same closed-loop bandwidth. To conclude, this work presents assessments of the non-minimum characteristic of the wireless power receivers and the proposed solution enables wireless power receivers to achieve better regulation.

TABLE II.
COMPARISON WITH DIFFERENT CONTROL METHODS FOR WIRELESS POWER RECEIVER

| | **This work** | [1] | [12] | [18] | [19] |
|---|---|---|---|---|---|
| Transmitter topology | **Full-bridge inverter** | Full-bridge inverter | Full-bridge inverter | Full-bridge inverter | Full-bridge inverter |
| Receiver topology | **Active rectifier + DC-DC converter** | Diode rectifier + Buck converter | Diode rectifier + Buck converter | Diode rectifier + Buck converter | Diode rectifier + Buck-boost converter |
| Topology for regulation | **Active rectifier** | Full bridge inverter and DC-DC converter | Buck converter | Buck converter | Buck-boost converter |
| Regulation objective | **Output voltage** | Output voltage | Output voltage | DC-link voltage | Output voltage |
| Control Method | **Receiver-side single-loop control** | Dual-side control | Receiver-side dual-loop control | Receiver-side model predictive control | Receiver-side sliding mode control |
| Number of sensors for control scheme | **1 voltage sensor** | 2 voltage sensors | 2 voltage sensors | 2 voltage sensors and 1 current sensor | 1 voltage sensors and 1 current sensor |
| Theoretical basis | **RHP zero elimination** | Transmitter-side coarse adjustment & Receiver-side fine regulation | Poles cancelation | Lyapunov stability criteria | Lyapunov stability criteria |
| Controller parameter design | **Model-based design** | N/A | Model-based design | Empirical | Routh-Hurwitz criteria |
| Wireless communication | **No** | Yes | No | No | No |
| RHP zero | **No** | Yes | Yes | N/A | N/A |
| Dynamic response | **Fast** | Slow | Medium | N/A | Medium |
| Stability Margin | **Wide** | N/A | Medium | Wide | Wide |
| Features | **High performance Minimum-phase system** | Distributed control burden | Ease of implementation | Large-signal stability | Large-signal stability |
| Disadvantage | **Extra component cost** | Significant delay due to wireless communication | Narrow operating range | High design complexity and computation burden | High design complexity and computation burden |